\documentclass[aps,prd,superscriptaddress,nofootinbib,preprint]{revtex4}

\usepackage{natbib}
\usepackage{ulem}
\usepackage{color}
\usepackage{graphicx}
\usepackage{multirow}
\usepackage{hyperref}
\usepackage{epstopdf}
\usepackage{amsmath,amssymb,amsthm,amsxtra,overpic,bbm,bm,float,epsfig}

\begin{document}

\title{Prospects for Pre-supernova Neutrino Observation in Future Large Liquid-scintillator Detectors} 

\author{Hui-Ling Li}
\email{lihuiling@ihep.ac.cn}
\affiliation{Institute of High Energy Physics, Chinese Academy of Sciences, Beijing 100049, China}

\author{Yu-Feng Li}
\email{liyufeng@ihep.ac.cn (corresponding author)}
\affiliation{Institute of High Energy Physics, Chinese Academy of Sciences, Beijing 100049, China}
\affiliation{School of Physical Sciences, University of Chinese Academy of Sciences, Beijing 100049, China}

\author{Liang-Jian Wen}
\email{wenlj@ihep.ac.cn (corresponding author)}
\affiliation{Institute of High Energy Physics, Chinese Academy of Sciences, Beijing 100049, China}

\author{Shun Zhou}
\email{zhoush@ihep.ac.cn}
\affiliation{Institute of High Energy Physics, Chinese Academy of Sciences, Beijing 100049, China}
\affiliation{School of Physical Sciences, University of Chinese Academy of Sciences, Beijing 100049, China}

\begin{abstract}
Before massive stars heavier than $(8 \cdots 10)$ solar masses evolve to the phase of a gravitational core collapse, they will emit a huge number of MeV-energy neutrinos that are mainly produced in the thermal processes and nuclear weak interactions. The detection of such pre-supernova (pre-SN) neutrinos could provide an important and independent early warning for the optical observations of core-collapse SNe. In this paper, we investigate the prospects of future large liquid-scintillator detectors for the observation of pre-SN neutrinos in both $\overline{\nu}^{}_e + p \to e^+ + n$ and $\nu (\overline{\nu}) + e^- \to \nu (\overline{\nu}) + e^-$ reaction channels, where $\nu$ ($\overline{\nu}$) denotes neutrinos (antineutrinos) of all three flavors. We propose a quantitative assessment of the capability in terms of three working criteria, namely, how far the SN distance can be covered, how long the early warning before the core collapse can be sent out, and how well the direction pointing to the SN can be determined. The dependence of the final results on the different models of progenitor stars, neutrino flavor conversions and the relevant backgrounds is also discussed.
\end{abstract}
\maketitle

\section{Introduction}

In the paradigm of the delayed neutrino-driven supernova (SN) explosions, massive stars heavier than $(8\cdots 10)$ solar masses will end their lives as core-collapse SNe by emitting a huge number of neutrinos of tens of MeV energies, which carry away most of the released gravitational binding energy of $3\times 10^{53}~{\rm erg}$~\cite{Woosley:2002zz, Janka:2017vlw}.  The successful detection of neutrinos from SN 1987A in Kamiokande-II~\cite{Hirata:1987hu}, IMB~\cite{Bionta:1987qt} and Baksan~\cite{Alekseev:1988gp} experiments has provided a strong support for this overall picture of core-collapse SNe.

Even prior to the core collapse, the evolved massive stars are experiencing significant neutrino emission, which is actually the most efficient mechanism of cooling after the phase of helium burning~\cite{Odrzywolek:2003vn, Odrzywolek:2010zz}. In the stellar environment, where both the matter density and the temperature may span quite a wide range, neutrinos are primarily produced in pairs by the plasma process $\gamma^* \to \nu + \overline{\nu}$, the photo process $\gamma + e^- \to e^- + \nu + \overline{\nu}$, the pair process $e^+ + e^- \to \nu + \overline{\nu}$ and the bremsstrahlung $e^- + Ze \to e^- + Ze + \nu + \overline{\nu}$, where $Ze$ stands for the heavy nuclei with the atomic number $Z$ and $\nu$ ($\overline{\nu}$) denotes neutrinos (antineutrinos) of all three flavors~\cite{Beaudet:1967zz, Munakata1985, Schinder:1986nh, Itoh:1989, Haft:1993jt, Itoh:1996, Guo:2016vls}. Only in the late stages of the stellar evolution, when the matter density and temperature are high enough, can neutrinos of MeV energies be generated in the stellar core and inner burning shells, which are called pre-SN neutrinos of particular interest for realistic detection. The contributions from nuclear weak interactions, including $e^{\pm}_{}$ capture on and $\beta^{\pm}_{}$ decays of heavy nuclei, to the pre-SN neutrino production become dominant in the last one hour or so prior to the core collapse~\cite{Yoshida:2016imf, Kato:2017ehj}. Therefore, the detection of pre-SN neutrinos can be implemented to diagnose the late-time evolution of massive stars and serves as the early-warning signal for the subsequent core-collapse SN explosions.

So far pre-SN neutrinos have not yet been detected. For the next nearby galactic SN, the detectability of pre-SN neutrinos in current and future neutrino detectors with different target materials has been investigated in several previous works, e.g., Refs.~\cite{Yoshida:2016imf, Kato:2017ehj, Odrzywolek:2003vn, Patton:2017neq}. Compared to the neutrino bursts from SN explosions, the typical energies of pre-SN neutrinos are much lower (i.e., a few MeV) and their fluxes are also much smaller, making their detection rather challenging. Three types of neutrino detectors have already been and will soon be available for this purpose. In the water-Cherenkov (wCh) detectors (e.g., Super-Kamiokande~\cite{Fukuda:2002uc, Abe:2016nxk} and Hyper-Kamiokande~\cite{Abe:2011ts, Abe:2018uyc}) and the liquid-scintillator (LS) detectors (e.g., KamLAND~\cite{Asakura:2015bga}, Borexino~\cite{Cadonati:2000kq}, JUNO~\cite{An:2015jdp} and LENA~\cite{Wurm:2011zn}), the inverse beta decay $\overline{\nu}^{}_e + p \to e^+ + n$ (IBD) and the elastic (anti)neutrino-electron scattering $\nu (\overline{\nu}) + e^- \to \nu (\overline{\nu}) + e^-$ ($e$ES) are two relevant detection channels. Recently, a dedicated study performed by the Super-Kamiokande collaboration~\cite{Simpson:2019xwo} shows that the pre-SN neutrinos can be successfully observed in the Gadolinium-doped detector of Super-Kamiokande via the IBD reaction, for which the delayed signal of $\gamma$ rays from the neutron capture on gadolinium will also be recorded. In addition, the liquid-argon time-projection chamber (LAr-TPC) detector of DUNE~\cite{Acciarri:2016crz} is particularly sensitive to electron neutrinos due to the charged-current interaction $\nu^{}_e + {^{40}\rm Ar} \to e^- + {^{40}\rm K}^*$. The relatively high energy threshold of this reaction restricts the experimental sensitivity to pre-SN neutrinos, since only the high-energy $\nu^{}_{e}$'s generated in the nuclear weak interactions in the last few minutes before collapse can be detected~\cite{Kato:2017ehj}.

In this work, we propose a quantitative assessment of the early-warning capability of future large LS detectors via the detection of pre-SN neutrinos~\footnote{We would like to remind that there is another preprint~\cite{Mukhopadhyay:2020ubs} appeared after our submission to the journal and arXiv, which is also devoted to the study of pre-supernova neutrinos and their directionality in large LS detectors.}. For illustration, the 20 kiloton LS detector of Jiangmen Underground Neutrino Observatory (JUNO)~\cite{An:2015jdp}, with a low threshold $E^{\rm th}_{\rm o} = 0.2~{\rm MeV}$ for the observable energy $E^{}_{\rm o}$ and an excellent energy resolution $3\%/\sqrt{E^{}_{\rm o}/(\rm MeV)}$, will be taken as an example. Different from the previous work~\cite{Guo:2019orq}, the present paper attempts to answer three questions regarding the early warning of core-collapse SNe: (1) how far the SN distance can be reached; (2) how long before the core collapse the warning can be sent out; (3) how well the direction pointing to the SN can be determined. In principle, we should take into account the neutrino events in both IBD and $e$ES channels, since the latter contributes remarkably to the total number of events and is sensitive to pre-SN neutrinos of all three flavors. The detection of pre-SN neutrinos other than $\overline{\nu}^{}_e$'s is obviously important to test the neutrino production mechanism in the late-time evolution of massive stars. In practice, however, the pre-SN neutrino events in the $e$ES channel show up as singles with only one visible signal in the detector and have very low energies, which will be easily contaminated by various backgrounds.
We stress that the high statistics of pre-SN neutrino events in future large LS detectors will greatly enhance the early warning capability for the nearby galactic SN, and open the possibility to reconstruct the direction pointing to the SN. The directional information together with the early warning will help to adjust the optical telescopes to catch the early-time light from the SN.

The remaining part of this paper is organized as follows. In Sec.~\ref{sec:II}, we describe the neutrino events in both IBD and $e$ES channels in a JUNO-like detector for different pre-SN neutrino models. The early-warning capability for the nearby galactic SNe is quantified in Sec.~\ref{sec:III} by considering only the IBD events and taking account of the background rate at JUNO as estimated in Ref.~\cite{An:2015jdp}. The dependence of the final results on the stellar models and neutrino flavor conversions is discussed. Finally, we summarize our main results and conclude in Sec.~\ref{sec:IV}.

\section{Pre-SN neutrino signals}\label{sec:II}

We start with the calculation of the event spectrum of pre-SN neutrinos. Since such a calculation was first carried out in Ref.~\cite{Odrzywolek:2003vn}, there have been more numerical stellar models of different progenitor star masses and the pre-SN neutrino data have been made available. In this work, the pre-SN neutrino models from Guo {\it et al.}~\cite{Guo:2019orq} and Kato {\it et al.}~\cite{Kato:2017ehj} will be adopted. The first set of Guo models contains the simulated neutrino data for four different progenitor star masses $\{12, 15, 20, 25\}~M^{}_{\odot}$ (in units of the solar mass $M^{}_{\odot}$), in which only the pair process $e^+ + e^- \to \nu + \overline{\nu}$ has been considered for the advanced stellar evolution. In second set of Kato models, the neutrino data for two progenitor star masses $\{12, 15\}~M^{}_{\odot}$ are given, where both thermal pair processes and nuclear weak interactions have been taken into account in the simulations. The luminosities of pre-SN neutrinos are computed by integrating the volume emissivity over the emission region, and thus should be functions of time and neutrino energy. To derive the fluxes of pre-SN neutrinos at the detectors, the luminosities will be divided by the average neutrino energies and further multiplied by the factor $1/D^{2}$ with $D$ being the distance to the SN. Notice that the luminosity of pre-SN neutrinos increases as the time approaches to the moment of core collapse, but it is a few orders of magnitude smaller than the typical luminosity $10^{52}~{\rm erg}\cdot {\rm s}^{-1}$ of the SN burst neutrinos~\cite{Janka:2017vlw}. The average energies of pre-SN neutrinos  are usually lower than $2$ MeV in most of the time before collapse, while they are rising quickly later such that the fraction of $\overline{\nu}^{}_{e}$'s with energies above the IBD threshold $E_{\nu}^{\rm th} = 1.8~{\rm MeV}$ becomes more and more significant. See, e.g., Refs.~\cite{Guo:2019orq} and \cite{Kato:2017ehj}, for more details about the luminosities and average energies of pre-SN neutrinos.

Then, the pre-SN neutrinos from the star will be reprocessed by flavor conversions before they arrive at the detectors. For these neutrinos, the refractive effects on the neutrino background and the resultant collective flavor oscillations will be ignored, since the total number density of neutrinos in the medium in the pre-SN phase is low~\cite{Pantaleone:1992eq, Samuel:1993uw, Duan:2010bg, Chakraborty:2016yeg, Mirizzi:2015eza}. Hence only the Mikheyev-Smirnov-Wolfenstein (MSW) matter effects on neutrino oscillations are relevant~\cite{Wolfenstein:1977ue, Mikheev:1986gs}. As demonstrated in Ref.~\cite{Guo:2019orq}, the flavor conversions with the MSW effects are highly adiabatic for the realistic matter density profile. Hence we follow the formalism of neutrino oscillations with the MSW effects in Ref.~\cite{Dighe:1999bi}, and calculate the fluxes of pre-SN neutrinos after flavor conversions. The final neutrino fluxes $F^{}_{\nu^{}_e}$ and $F^{}_{\nu^{}_x}$ can be expressed in terms of the initial ones $F^0_{\nu^{}_e}$ and $F^0_{\nu^{}_x}$, namely, $F^{}_{\nu^{}_e} = p F^0_{\nu^{}_e} + (1 - p) F^0_{\nu^{}_x}$ and $F^{}_{\nu^{}_x} = 0.5 (1 - p) F^0_{\nu^{}_e} + 0.5 (1 + p) F^0_{\nu^{}_x}$, where $x$ refers to the muon or tau flavor. Similar formulas exist for antineutrinos fluxes $F^{}_{\overline{\nu}^{}_e}$ and $F^{}_{\overline{\nu}^{}_x}$ and the initial ones $F^0_{\overline{\nu}^{}_e}$ and $F^0_{\overline{\nu}^{}_x}$, where we need only to replace the survival probability $p$ for neutrinos in the aforementioned formulas by that $\overline{p}$ for antineutrinos. In the case of normal neutrino mass ordering (NO) with $m^{}_1 < m^{}_2 < m^{}_3$, where $m^{}_i$ (for $i = 1, 2, 3$) are absolute neutrino masses, we have $p = \sin^2 \theta^{}_{13} \approx 0.022$ and $\overline{p} = \cos^2 \theta^{}_{13} \cos^2 \theta^{}_{12} \approx 0.687$. In the case of inverted neutrino mass ordering (IO) with $m^{}_3 < m^{}_1 < m^{}_2$, we obtain $p = \sin^2\theta^{}_{12} \cos^2\theta^{}_{13} \approx 0.291$ and $\overline{p} = \sin^2 \theta^{}_{13} \approx 0.022$. Here the best-fit values of two relevant neutrino mixing angles from the latest global analysis of neutrino oscillation data have been used~\cite{Tanabashi:2018oca}. In addition, we have neglected the terrestrial matter effects~\cite{Dighe:2003jg, Mirizzi:2006xx, Borriello:2012zc, Liao:2016uis} for pre-SN neutrinos, where the day-night effects are estimated be much smaller than $1\%$ for the pre-SN neutrinos with the average energy of around 2 MeV compared to the $3\%$ day-night asymmetry for $^{8}{\rm B}$ solar neutrinos observed in Super-Kamiokande~\cite{Abe:2016nxk}.

\begin{figure}[!t]
\begin{center}
\begin{tabular}{l}
\hspace{-0.6cm}
\includegraphics[width=0.59\textwidth]{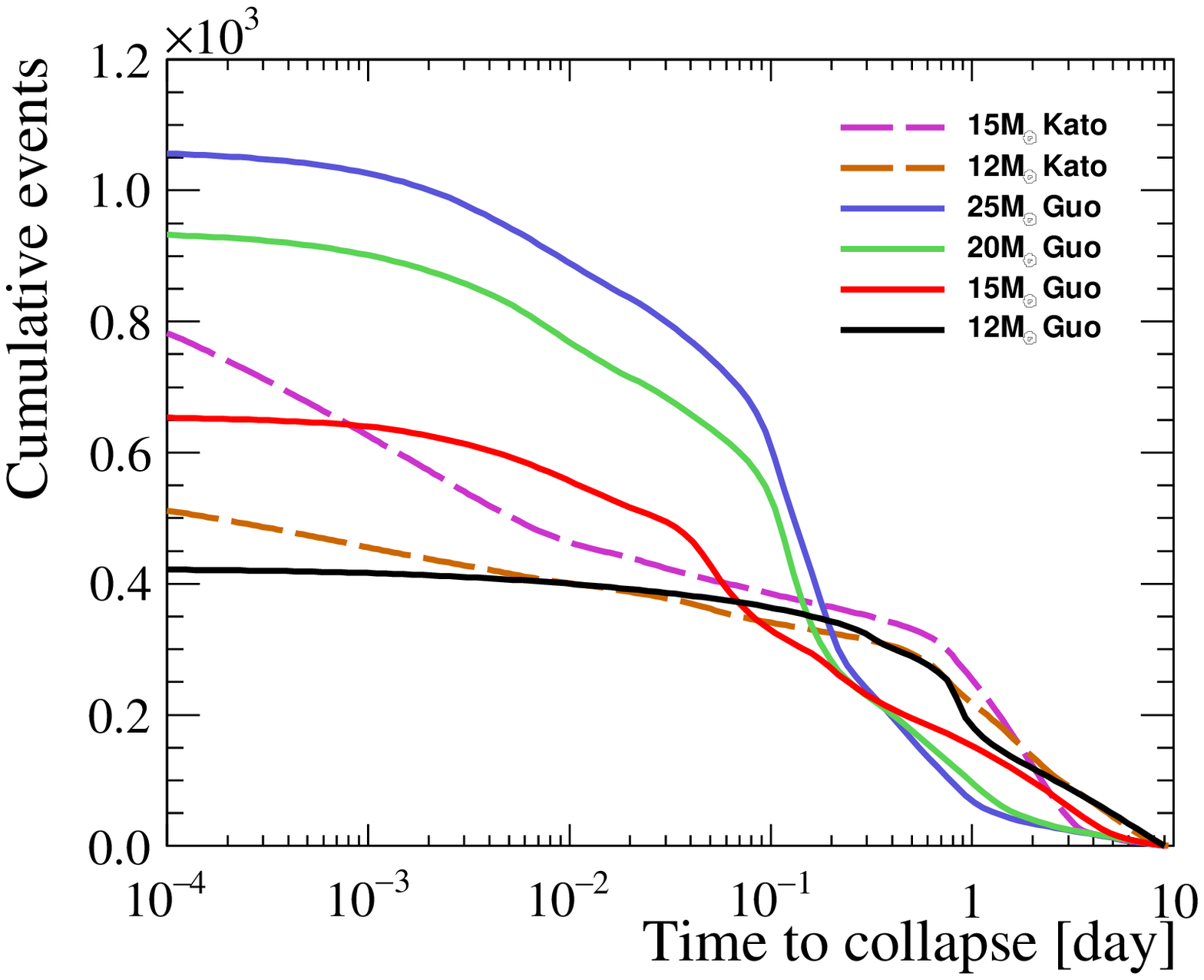}
\hspace{-1.5cm}
\includegraphics[width=0.59\textwidth]{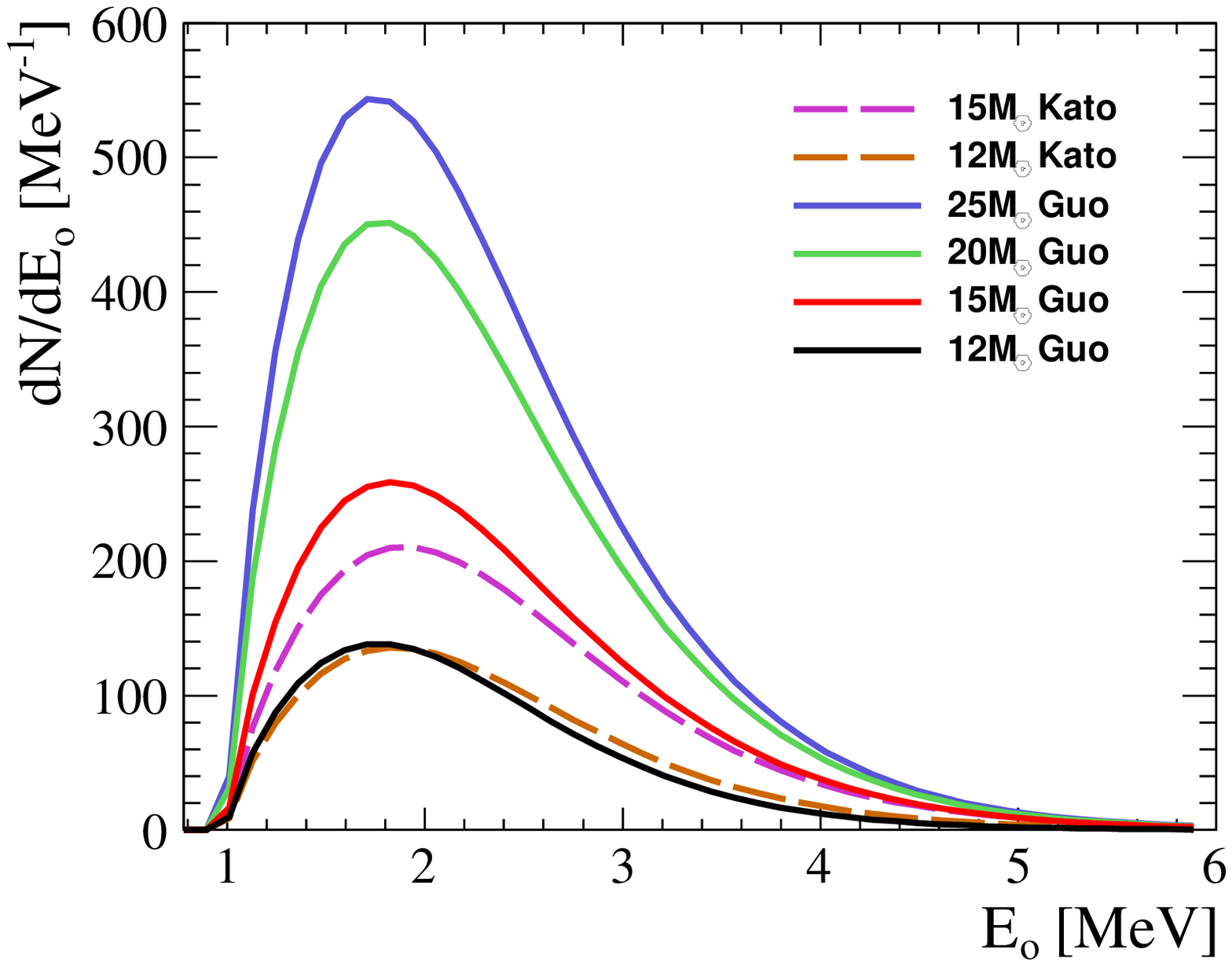}
\\
\hspace{-0.6cm}
\includegraphics[width=0.59\textwidth]{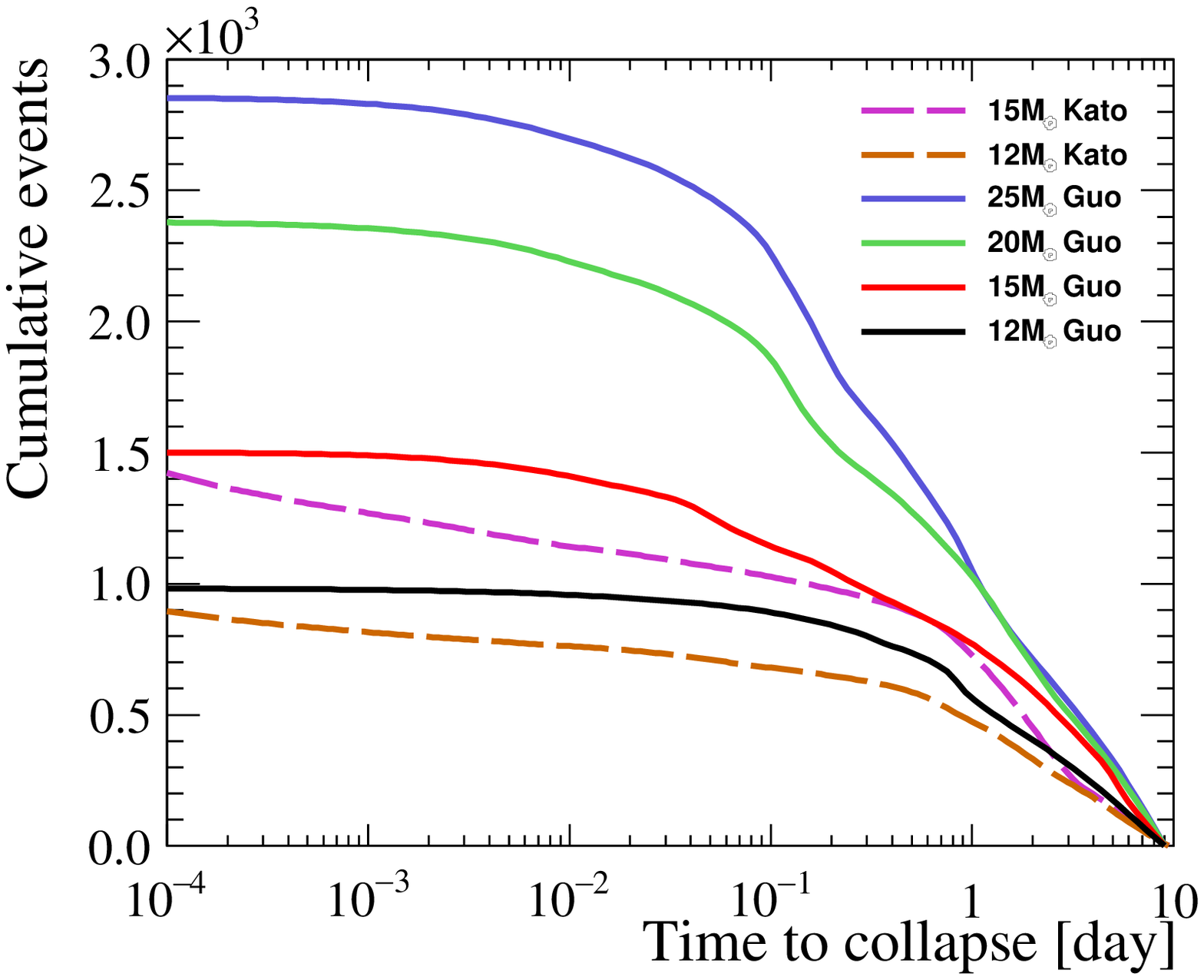}
\hspace{-1.5cm}
\includegraphics[width=0.59\textwidth]{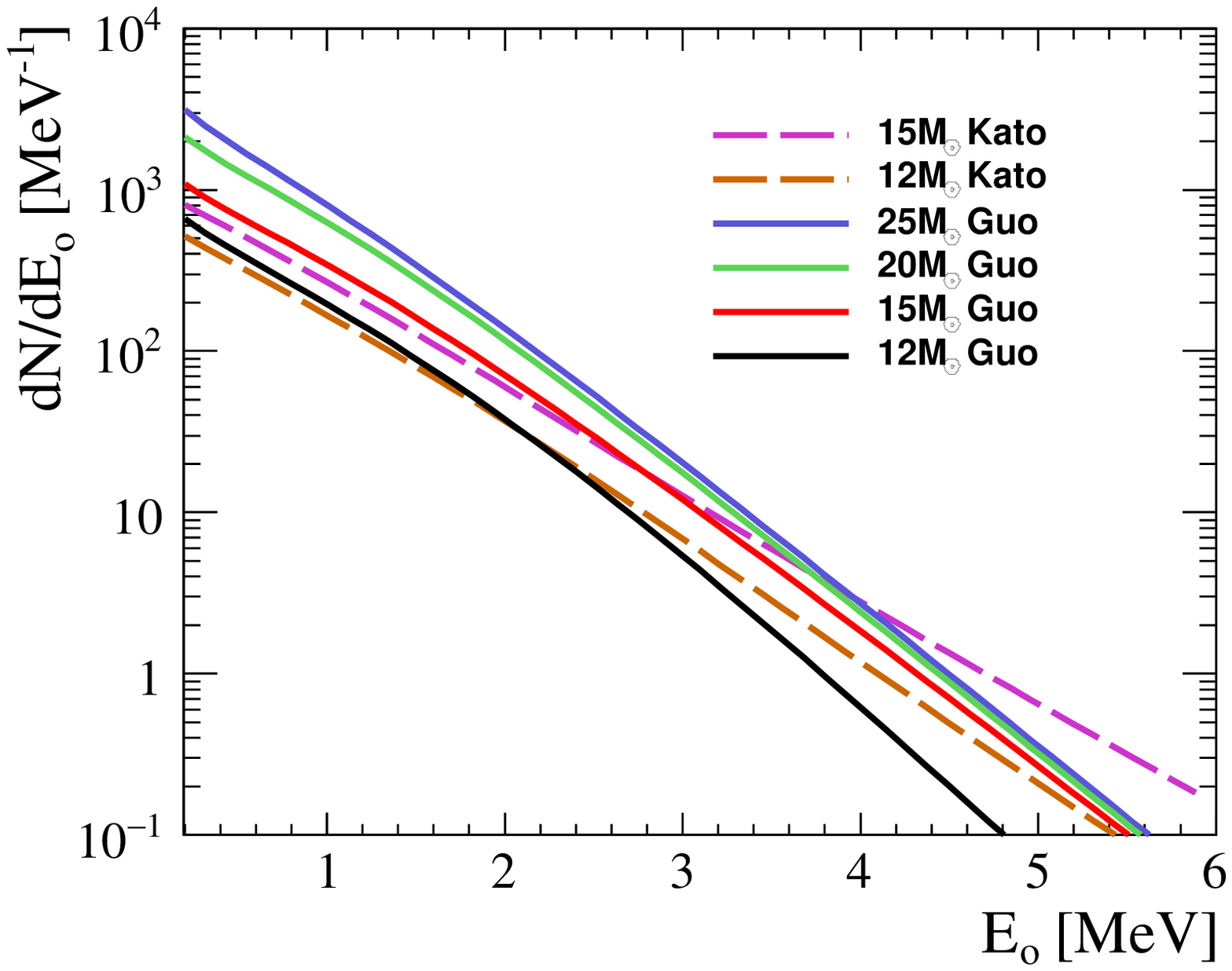}
\end{tabular}
\end{center}
\vspace{-0.6cm}
\caption{In the first row, the cumulative distribution of IBD events with respect to the time before the collapse (left panel) and the IBD event spectrum versus the observable energy (right panel) are given. In the second row, the same results are shown for the $e$ES events. Notice that the event spectrum has been obtained by integrating the event rate over the last one day prior to collapse, while assuming a pre-SN star at $D = 0.2~{\rm kpc}$ and the normal neutrino mass ordering.
\label{fig:evtIBDeES}}
\end{figure}

Finally, we take the JUNO LS detector as an example to compute the event spectrum. The JUNO detector utilizes 20 kiloton LS, of which 12$\%$ are protons and 88$\%$ are carbon atoms, and the energy resolution is expected to be $3\%/\sqrt{E^{}_{\rm o}/{\rm (MeV)}}$~\cite{An:2015jdp}. For the SN burst neutrinos, the event rates of all three flavor neutrinos and antineutrinos in the LS detectors have been systematically studied in Refs.~\cite{Lujan-Peschard:2014lta, Lu:2016ipr}. For the pre-SN neutrinos, whose energies are much lower, it is impossible to detect them via the neutrino-proton elastic scattering, since the recoil energies of the final-state protons are far below the threshold $E^{\rm th}_{\rm o} = 0.2~{\rm MeV}$ for the observable energies in JUNO. In addition, both the charged- and neutral-current interactions of neutrinos with the carbon nuclei are absent due to the low energies of pre-SN neutrinos. Therefore, only the IBD and $e$ES channels are available. Given the neutrino fluxes and a JUNO-like LS detector, the explicit calculations of the event rates in the IBD and $e$ES channels can be found in Ref.~\cite{Lu:2016ipr, Li:2017dbg}, and our final numerical results are shown in Fig.~\ref{fig:evtIBDeES} and Table~\ref{table:EventsIBDeES}. Some comments on the results are in order.
\begin{table}[!t]
\setlength{\tabcolsep}{11pt}
\centering
\begin{tabular}{ccccccccc}
\hline
\multicolumn{1}{c}{\multirow{2}{*}{Model}}  & \multicolumn{1}{c}{}   & \multicolumn{3}{c}{IBD } & \multicolumn{1}{c}{}  & \multicolumn{3}{c}{ $e$ES } \\
\cline{3-5}\cline{7-9} \multicolumn{1}{c}{}  & \multicolumn{1}{c}{}  & \multicolumn{1}{c}{NO} & \multicolumn{1}{c}{} & \multicolumn{1}{c}{IO} & \multicolumn{1}{c}{}  & \multicolumn{1}{c}{NO} & \multicolumn{1}{c}{} & \multicolumn{1}{c}{IO}  \\
\hline
\multicolumn{1}{l}{$12 M_{\odot}$ Guo} & \multicolumn{1}{c}{}  & \multicolumn{1}{c}{240 (432)} & \multicolumn{1}{c}{} & \multicolumn{1}{c}{74 (129)} & \multicolumn{1}{c}{}  & \multicolumn{1}{c}{420 (1017)} & \multicolumn{1}{c}{} & \multicolumn{1}{c}{475 (1156)} \\
\multicolumn{1}{l}{$15 M_{\odot}$ Guo} & \multicolumn{1}{c}{}  & \multicolumn{1}{c}{503 (659)} & \multicolumn{1}{c}{} & \multicolumn{1}{c}{152 (195)} & \multicolumn{1}{c}{} &  \multicolumn{1}{c}{733 (1549)} & \multicolumn{1}{c}{} & \multicolumn{1}{c}{824 (1764)} \\
\multicolumn{1}{l}{$20 M_{\odot}$ Guo} & \multicolumn{1}{c}{}   & \multicolumn{1}{c}{842 (941)} & \multicolumn{1}{c}{} & \multicolumn{1}{c}{245 (270)} & \multicolumn{1}{c}{}  & \multicolumn{1}{c}{1355 (2447)} & \multicolumn{1}{c}{} & \multicolumn{1}{c}{1519 (2787)} \\
\multicolumn{1}{l}{$25 M_{\odot}$ Guo} & \multicolumn{1}{c}{} & \multicolumn{1}{c}{993 (1064)} & \multicolumn{1}{c}{} & \multicolumn{1}{c}{287 (305)} & \multicolumn{1}{c}{}   & \multicolumn{1}{c}{1807 (2926)} & \multicolumn{1}{c}{} & \multicolumn{1}{c}{2033 (3339)} \\
\multicolumn{1}{l}{$12 M_{\odot}$ Kato} & \multicolumn{1}{c}{} & \multicolumn{1}{c}{312 (531)} & \multicolumn{1}{c}{} & \multicolumn{1}{c}{84 (151)} & \multicolumn{1}{c}{}   & \multicolumn{1}{c}{466 (952)} & \multicolumn{1}{c}{} & \multicolumn{1}{c}{614 (1140)} \\
\multicolumn{1}{l}{$15 M_{\odot}$ Kato} & \multicolumn{1}{c}{}  & \multicolumn{1}{c}{578 (833)} & \multicolumn{1}{c}{} & \multicolumn{1}{c}{122 (197)} & \multicolumn{1}{c}{}  & \multicolumn{1}{c}{775 (1517)} & \multicolumn{1}{c}{} & \multicolumn{1}{c}{1043 (1888)} \\

\hline
\end{tabular}
\vspace{0.1cm}
\caption{The integrated number of pre-SN neutrino events in both IBD and $e$ES channels at a JUNO-like detector over the last one day before collapse, where the pre-SN star at $D = 0.2~{\rm kpc}$ is assumed. For comparison, the event numbers in the last ten days before collapse have been given in parentheses.}
\label{table:EventsIBDeES}
\end{table}
\begin{itemize}
\item In the upper row of Fig.~\ref{fig:evtIBDeES}, we show the cumulative distribution of the IBD events with respect to the time (in units of days) before the collapse in the left panel, and the IBD event spectrum versus the observable energy in the right panel. In both panels, the results for the Guo models with four different progenitor masses and the Kato models with two different progenitor masses are represented by the solid and dashed curves, respectively. Notice that the event spectrum has been obtained by integrating the event rate over the last one day prior to collapse.  The distance of the progenitor stars in our calculations is taken to be $D = 0.2~{\rm kpc}$, and neutrino flavor conversions with the MSW effects in the NO case are assumed. Several interesting observations can be made. First, a larger number of pre-SN neutrino events is generally expected for a heavier progenitor mass. Second, neutrinos in the last one day make a dominant contribution to the total IBD events. Third, the IBD events are concentrated on the observable energies around $E^{}_{\rm o} \approx 2~{\rm MeV}$ for all the stellar models. Although neutrino events with higher observable energies exist, their contribution will be suppressed. As the energy threshold of Super-Kamiokande is about $E^{\rm th}_{\rm o} \approx 3.5~{\rm MeV}$~\cite{Abe:2016nxk}, most of the IBD events will not be recognized by the detector even with Gd doped.
\item In the bottom row of Fig.~\ref{fig:evtIBDeES}, similar results in the $e$ES channel have been given, where the input parameters are the same as those for the IBD channel. Although the cross section of the $e$ES is much lower than that of the IBD, more $e$ES events are obtained for two reasons. First, all neutrinos and antineutrinos of three flavors contribute to the $e$ES events. Second, the energy threshold for the IBD reaction is $E^{\rm th}_{\nu} = 1.8~{\rm MeV}$ such that quite a number of low-energy neutrinos cannot be detected in the IBD channel. As one can observe from the right panel of the bottom row of Fig.~\ref{fig:evtIBDeES}, most $e$ES events appear in the region below $E^{}_{\rm o} < 2~{\rm MeV}$. However, it should be noted that the low-energy $e$ES events will be recognized as singles and contaminated by a high rate of background noises induced by the radioactivity of the detector materials and cosmogenic isotopes. Hence a very careful analysis of background reduction needs to be carried out. The IBD events will be clearly distinguished from the backgrounds with the help of the coincidence between the prompt signal of the positron annihilation and the delayed signal of neutron capture on hydrogen.
\end{itemize}

Compared to the Guo models, the cumulative events of IBD and $e$ES channels increase within the last few minutes before collapse in the Kato models, which can be attributed to the extra nuclear weak interactions. As this difference leads to more high-energy pre-SN neutrinos in the Kato models, one can see from the right panel of the bottom row of Fig.~\ref{fig:evtIBDeES} that more energetic $e$ES events show up for the progenitor masses $\{12, 15\}~M^{}_{\odot}$. In Table~\ref{table:EventsIBDeES}, we summarize the total number of IBD and $e$ES events in the last one day before collapse for a pre-SN star at $D = 0.2~{\rm kpc}$, while the number in the last ten days are given in parentheses. For the progenitor mass of $25~M_{\odot}$, there will be more than 1000 IBD events and 3000 $e$ES events registered in a JUNO-like LS detector. In the IBD channel, the event number depends crucially on the neutrino mass ordering. For instance, compared to the NO case, the event number can be suppressed by a factor of three or four in the IO case. In contrast, the number of events in the $e$ES channel for IO is slightly higher than that for NO. Therefore, as pointed out in Ref.~\cite{Guo:2019orq}, the observations in both IBD and $e$ES channels could help pin down neutrino mass ordering in a model-independent way.

\section{Pre-SN early warning}\label{sec:III}

With the signals of pre-SN neutrinos given in the previous section, now we examine how these signals can be implemented as an early warning for a multi-messenger observations of core-collapse SNe, in particular for the optical observation in the very early time. Given the clean coincidence signals of IBD events, one can monitor the IBD event candidates within a given sliding time window to determine whether a pre-SN early warning should be sent out. In this section, we explore the sensitivity to a pre-SN star as well as the pre-SN pointing capability by using the IBD events in a JUNO-like LS detector.

\subsection{Backgrounds}
\begin{figure}[!t]
\begin{center}
\begin{tabular}{l}
\includegraphics[scale=0.7]{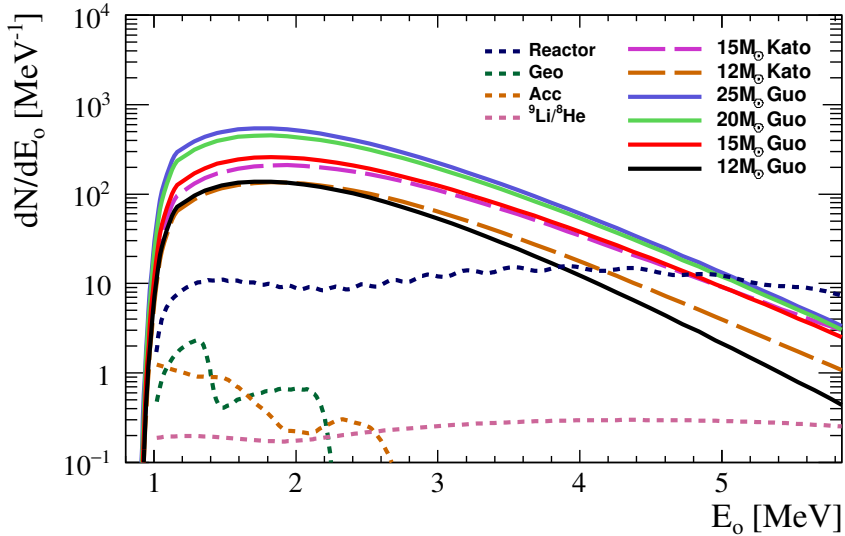}
\end{tabular}
\end{center}
\vspace{-0.5cm}
\caption{ The spectrum of the IBD events (solid and long dashed curves) in the last one day before collapse, where the pre-SN star at $D = 0.2 ~{\rm kpc}$ and neutrino oscillations in the NO case are assumed. The background events are represented by short dashed curves, including those from reactor neutrinos with the total thermal power of 36 ${\rm GW^{}_{th}}$ at a distance of 53 km (dark blue), geo-neutrinos (olive), accidental coincidence signals (brown) and $\beta$-$n$ decays of cosmogenic $^{9}{\rm Li}/^{8}{\rm He}$ (pink) within one day in JUNO~\cite{An:2015jdp}.
\label{fig:SigBkg}}
\end{figure}
Since the luminosities and energies of pre-SN neutrinos are relatively low, the level of background noises for the neutrino signals in the LS detectors will be critically important. Therefore, we first specify the backgrounds that are used in our later discussions. As the JUNO detector is taken for illustration, we collect all the background information about the JUNO detector, which are publicly obtainable from Ref.~\cite{An:2015jdp}. At JUNO, the sources of backgrounds for the IBD events of pre-SN $\overline{\nu}^{}_{e}$'s arise primarily from the reactor neutrinos and sub-dominantly from the geo-neutrinos, accidental coincidence signals and $\beta$-$n$ decays of cosmogenic $^{9}{\rm Li}/^{8}{\rm He}$.

In Fig.~\ref{fig:SigBkg}, the spectrum of IBD events from pre-SN neutrinos in the last one day prior to collapse has been depicted for different stellar models, where the observable energy spectra of relevant background events from Ref.~\cite{An:2015jdp} within one day are also shown for comparison. In the calculation of IBD events, we assume the pre-SN star at $0.2~{\rm kpc}$ and neutrino oscillations with the MSW effects in the NO case. The IBD selection efficiency can be approximately taken to be 73$\%$, as explained in Ref.~\cite{An:2015jdp}. To maximize the signal-to-background ratio while retaining more than $90\%$ of pre-SN IBD events, we choose the energy window as $0.9~{\rm MeV} \le E^{}_{\rm o} \le 3.5$ MeV to evaluate the experimental sensitivity in the following subsection. Within such an energy window, the total background rate is almost uniformly distributed and approximates to 28 events per day, as indicated in Fig.~\ref{fig:SigBkg}, including 25.3 events from reactor neutrinos, 1.1 events from geo-neutrinos, 0.9 events from accidental coincidence signals and 0.5 events from cosmogenic $^{9}{\rm Li}/^{8}_{}{\rm He}_{}$ decays.

\subsection{Sensitivity}

In order to quantitatively assess the experimental sensitivity to the pre-SN neutrinos, we implement the model-independent Poisson likelihood ratio and define the statistical significance as follows
\begin{eqnarray}
S = \sqrt{ 2n^{}_{\rm o}{\rm ln}\left(1+s/b \right)-2s} \; ,
\label{eq:Significance}
\end{eqnarray}
where $S$ is given in units of the standard deviation $\sigma$ and $n^{}_{\rm o}$ denotes the total number of observed events with $s$ being the number of pre-SN $\bar{\nu}^{}_{e}$ events and $b$ being the background events. With the significance introduced in Eq.~(\ref{eq:Significance}), we examine the following two questions: (1) how far the SN distance can be reached if a $3\sigma$ significance is required to report the detection of pre-SN neutrinos within the last one day prior to collapse; (2) given the pre-SN star at $D = 0.2~{\rm kpc}$, how long before the collapse the early warning can be sent out with a $3\sigma$ significance for the detection of pre-SN neutrinos.

\begin{figure}[!t]
\begin{center}
\begin{tabular}{l}
\includegraphics[scale=0.7]{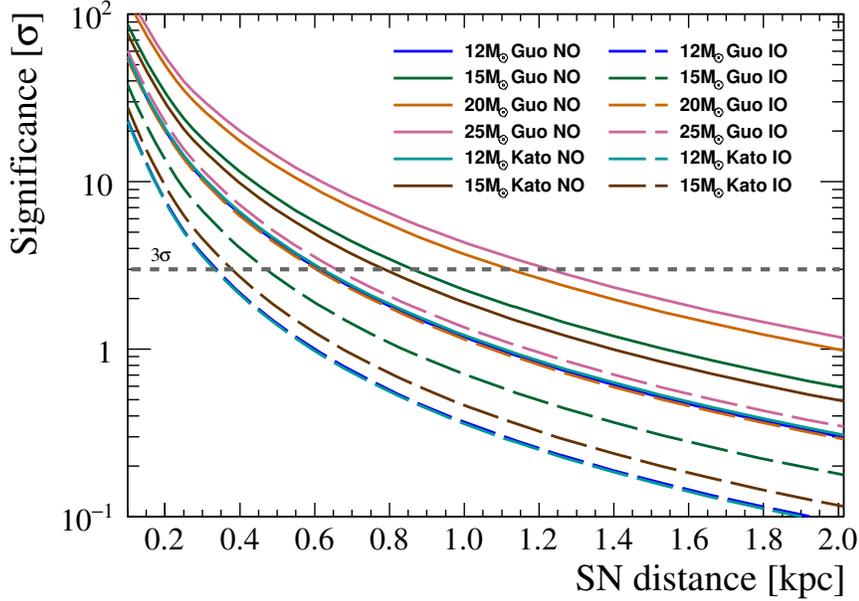}
\end{tabular}
\end{center}
\vspace{-0.5cm}
\caption{The expected significance of pre-SN neutrino induced IBD events in the last one day before collapse as a function of the pre-SN distance. A JUNO-like detector with 20 kiloton LS is assumed and neutrino oscillations with the MSW effects with NO (solid curves) or IO (long dashed curves) are considered.
\label{fig:sigDist}}
\end{figure}

Regarding the above questions, we have tested different choices of the sliding time window for observations, ranging from 0.1 days to 4 days, and finally fix it at one day in this work to make a balance between the early-warning time and the capability of reaching possible SN candidates within our galaxy. Such one day time window covers IBD event candidates spanning from an arbitrary given time to its last one day period, which is independent of the SN time frame. Given 28 background events within one day, one can calculate the corresponding detection significance by using Eq.~(\ref{eq:Significance}), which is actually a function of the pre-SN distance, as shown in Fig.~\ref{fig:sigDist}. Similarly, one can plot this function for different progenitor masses and in both NO and IO cases. As we have already explained, the IBD events of pre-SN neutrinos are obtained by integrating the rate over the last one day before collapse within the energy window $0.9~{\rm MeV} \le E^{}_{\rm o} \le 3.5~{\rm MeV}$ and by applying the detection efficiency of $73\%$. The event number is inversely proportional to the square of the distance to the SN. If a $3\sigma$ significance, shown as the horizontal dashed line in Fig.~\ref{fig:sigDist}, is required to send out the warning of a pre-SN, a JUNO-like LS detector will be sensitive to a 25 $M^{}_{\odot}$ progenitor star in the Guo models up to $1.23~{\rm kpc}$ in the NO case. In the worst situation, it is still sensitive to a 12 $M^{}_{\odot}$ star in the Kato models up to $0.34~{\rm kpc}$ in the IO case. Some general conclusions can be made. First, the significance will be higher for heavier masses of progenitor stars, as the number of IBD events will be larger. Second, the significance becomes worse in the IO case, compared to the result in the NO case. In any case, the sensitivity to the pre-SN distance in the future large LS detectors is much better than that in the LS detector of KamLAND, e.g $0.69~{\rm kpc}$ at a $3\sigma$ significance for a 25 $M^{}_{\odot}$ star in the NO case~\cite{Asakura:2015bga}, and the Gd-doped wCh detector of Super-Kamiokande, e.g 0.4 kpc for a 25 $M^{}_{\odot}$ star in the NO case~\cite{Asakura:2015bga}.

\begin{figure}[!t]
\begin{center}
\begin{tabular}{l}
\includegraphics[scale=0.7]{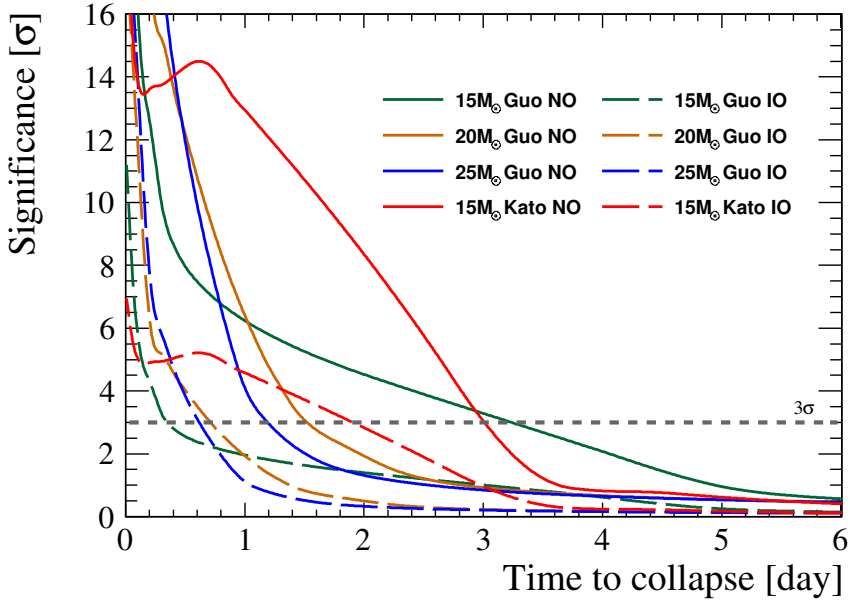}
\end{tabular}
\end{center}
\vspace{-0.8cm}
\caption{The expected significance of the pre-SN neutrino induced IBD events as a function of the time prior to collapse, where the SN distance $D = 0.2~{\rm kpc}$ as for the Betelgeuse is assumed. A JUNO-like detector is taken and neutrino oscillations with the MSW effects in the NO case (solid curves) and in the IO case (long dash lines) are considered.
\label{fig:sigTime}}
\end{figure}

As is well known, the red supergiant, Betelgeuse is one of the possible nearby galactic SN candidates with the uncertain mass and distance~\cite{Harper:2008,Smith:2008ef,Neilson:2011ta, Dolan:2017}, e.g its distance is determined as $D = 197 \pm 45 $ pc and its mass is within $M = (17\cdots 25)~M^{}_{\odot}$ in Ref.~\cite{Harper:2008}. Taking the SN distance at $D = 0.2~{\rm kpc}$ as for Betelgeuse, we explore how long before the core collapse the early-warning message can be sent out when requiring the $3\sigma$ significance of the detection of pre-SN neutrinos. The pre-SN neutrinos from the Guo models of the progenitor masses $\{15, 20, 25\}~M^{}_{\odot}$ and from the Kato model of $15~M^{}_{\odot}$ are taken into account, together with neutrino oscillations in both NO and IO cases. In Fig.~\ref{fig:sigTime}, the significance of pre-SN IBD events is plotted as a function of the time before collapse, where a JUNO-like 20 kiloton LS detector is assumed and different pre-SN neutrino models together with NO or IO are considered. The expected early-warning time before collapse for a possible SN candidate at $D = 0.2~{\rm kpc}$ has been extracted from Fig.~\ref{fig:sigTime} and listed in Table~\ref{table:sigTime}, where the $3\sigma$ significance of the detection of pre-SN neutrinos is required and a JUNO-like detector is assumed. In the optimal scenario, the pre-SN early warning could be sent out about 3.2 days before collapse in the NO case and 0.3 days in the IO case for the Guo model with the progenitor mass of 15 $M^{}_{\odot}$. In any case, a JUNO-like LS detector will be able to send out an early warning for the Betelgeuse-like star at least eight hours before collapse.

Finally let us comment on the issue of the false alert rate in the sliding window method, which depends on the refreshing time length and the number of independent trials during the monitoring process. For instance, for a one-hour refreshing time in the one-day sliding window, the average false alert rate induced by the background fluctuation is less than 1 per month with the $3\sigma$ significance criteria, which is validated with 1000 one-year trials in a toy Monte Carlo simulation.
In practice, the sliding window duration and the refreshing time length should be optimized according to experimental requirements with the balanced observation significance and false alert rate.

\begin{table}[!t]
\setlength{\tabcolsep}{11pt}
\centering
\begin{tabular}{cccccc}
\hline
\multicolumn{1}{c}{\multirow{2}{*}{Mass Ordering}}  & \multicolumn{1}{c}{} & \multicolumn{4}{c}{Model} \\
\cline{3-6}\multicolumn{1}{c}{}  & \multicolumn{1}{c}{}  & \multicolumn{1}{c}{15 $M_{\odot}$ Guo} & \multicolumn{1}{c}{20 $M_{\odot}$ Guo} & \multicolumn{1}{c}{25 $M_{\odot}$ Guo} & \multicolumn{1}{c}{15 $M_{\odot}$ Kato} \\
\hline
\multicolumn{1}{l}{NO} &\multicolumn{1}{c}{}  & \multicolumn{1}{c}{3.2}  & \multicolumn{1}{c}{1.5} & \multicolumn{1}{c}{1.2} & \multicolumn{1}{c}{3.0} \\
\multicolumn{1}{l}{IO}  &\multicolumn{1}{c}{}  & \multicolumn{1}{c}{0.3}  & \multicolumn{1}{c}{0.7} & \multicolumn{1}{c}{0.6} & \multicolumn{1}{c}{1.9}  \\
\hline
\end{tabular}
\vspace{0.3cm}
\caption{The expected time (in units of days) before collapse for a possible SN candidate at $D = 0.2~{\rm kpc}$ to reach a $3\sigma$ significance of the detection of pre-SN neutrinos in a 20 kiloton LS detector, where different pre-SN neutrino models and neutrino oscillations in either NO or IO case are considered. }
\label{table:sigTime}
\end{table}

\subsection{Pre-SN pointing}
\begin{figure}[!t]
\begin{center}
\begin{tabular}{l}

\includegraphics[scale=0.9]{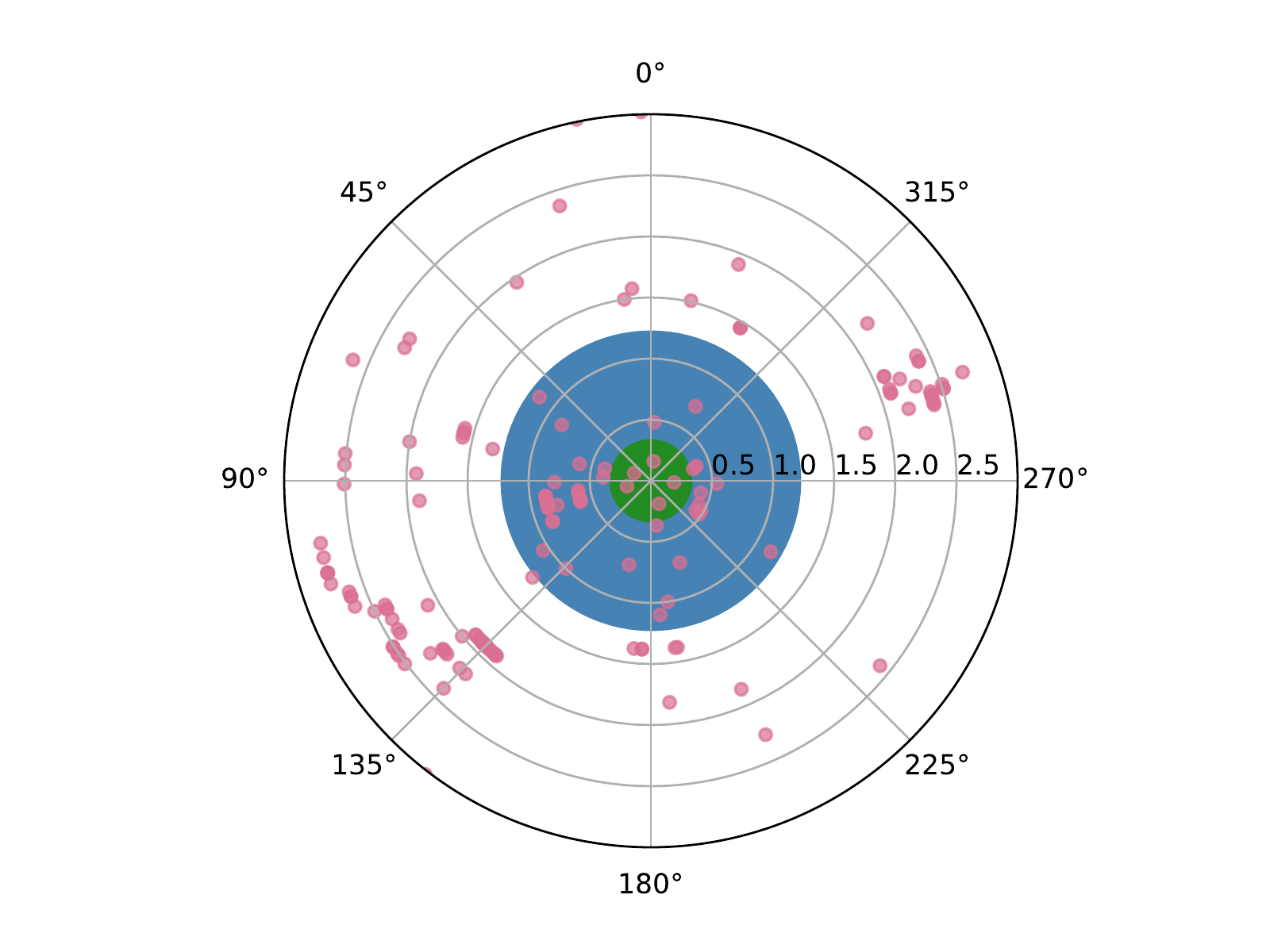}
\end{tabular}
\end{center}
\vspace{-0.7cm}
\caption{The nearby red supergiant (RSG) stars are plotted in the galactic coordinate system with the galactic longitude and radius (in units of kpc). For the adopted pre-SN neutrino models, a JUNO-like LS detector is sensitive to 45 RSG stars in the blue region in the optimal scenario (for $D < 1.23$ kpc), and to 5 RSG stars in the green region in the worst-case scenario (for $D < 0.34$ kpc), where a $3\sigma$ significance is required for the detection of pre-SN neutrinos.
\label{fig:sigRSG}}
\end{figure}
As we have mentioned, a JUNO-like LS detector is sensitive to the galactic pre-SN stars up to the distance of $(0.34 \cdots 1.23)~{\rm kpc}$, depending on the stellar models and neutrino mass ordering. According to the collection of red supergiant (RSG) stars in our galaxy from Ref.~\cite{Nakamura:2016kkl}, there are 45 RSG stars within 1.23 kpc and 5 within 0.34 kpc, which have been shown respectively in the blue and green regions in Fig.~\ref{fig:sigRSG}. Here the galactic coordinate system with the galactic longitude (in units of degrees) and radius (in units of kpc). The detailed information about 5 RSG stars within 0.34 kpc are quoted from Ref.~\cite{Nakamura:2016kkl}: Enif (150 pc), Antares (160 pc), Suhail (190 pc), Betelgeuse (200 pc),  and Zeta Cephei(200 pc). Since the RSG stars may turn into core-collapse SNe, the pre-SN early warning can be sent to other communities via the SuperNova Early Warning System (SNEWS)~\cite{Antonioli:2004zb}, which subsequently informs gravitational-wave observatories and astronomers to prepare for the upcoming SN explosion. In this sense, the pre-SN direction is of great importance for astronomers to adjust their telescopes to observe the early optical signals from the corresponding SN candidate. Unlike the case of SN burst neutrinos, the wCh detectors can hardly provide the directional information by detecting less energetic pre-SN neutrinos due to its relatively high energy threshold. Keeping this point in mind, we explore the pre-SN pointing capability of future large LS detectors.

\begin{figure}[!t]
\begin{center}
\begin{tabular}{c}
\vspace{-0.5cm}
\includegraphics[scale=0.6]{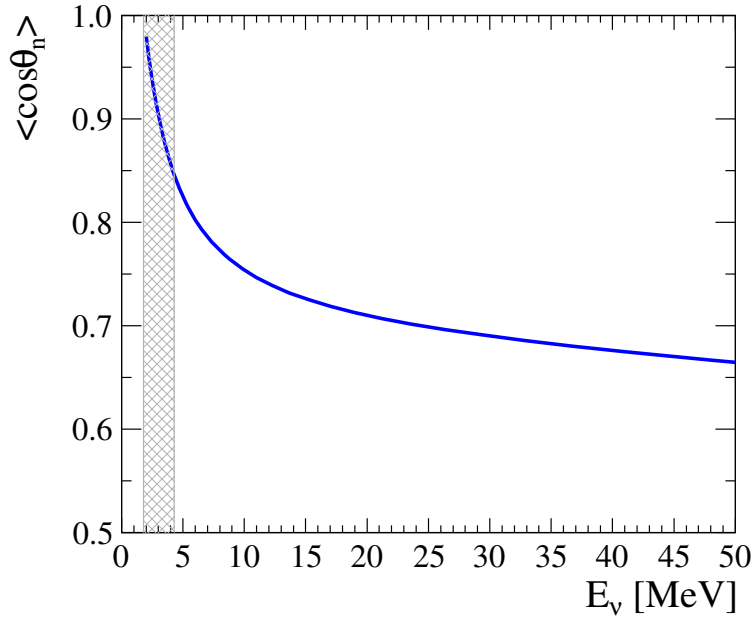}
\end{tabular}
\end{center}
\vspace{-0.7cm}
\caption{The average neutron direction of IBD events versus the energy of incident $\overline{\nu}_{e}$ is shown as the blue line. The shade refers to the energy range between 1.8 MeV and 4.3 MeV for pre-SN neutrinos. Here $\theta_{n}$ is defined as the angle between $\overline{\nu}_{e}$ and the initial neutron direction and $\left< \cos{\rm \theta_{n}} \right>$ is calculated with the IBD cross section of Ref.~\cite{Strumia:2003zx}. 
\label{fig:neutronCos}}
\end{figure}

The main strategy for the LS detector to determine the SN direction is to make use of the anisotropy of the IBD signals, as first discussed in Refs.~\cite{Vogel:1999zy, Apollonio:1999jg, Fischer:2015oma}. In our calculations, we implement the simulations with Geant4 to obtain the information about energy deposition of charged particles and photons in the LS detector. First, the pre-SN $\overline{\nu}^{}_{e}$'s follow the energy distribution as predicted by the Guo model for a progenitor mass of 25 $M^{}_{\odot}$ and the IBD events are then uniformly distributed in a JUNO-like LS detector. The LS detector identifies the IBD events from the pre-SN $\overline{\nu}^{}_{e}$'s as coincidence signals from the prompt $e^{+}_{}$ annihilation and the delayed signal of neutron capture on hydrogen. The prompt $e^{+}_{}$ is produced with a tiny backward anisotropy~\cite{Vogel:1999zy} and subsequently annihilates with electrons after a flight of $0.5$ cm or so, in which two 0.511 MeV $\gamma$ rays are isotropically emitted in opposite directions. The energy-weighted mean position of the prompt signal is about 6.5 cm away to the production point. Considering the vertex resolution of positrons, i.e, 7 cm at 1 MeV without the transit time spread of PMT considered from Ref.~\cite{Liu:2018fpq}, it is reasonable to assume that the annihilation position or the mean energy-deposition position is approximately the point where $\overline{\nu}^{}_{e}$ interacts with protons. On the other hand, the neutron from the IBD reaction is emitted in the forward hemisphere with respect to the incident $\overline{\nu}^{}_{e}$ direction. This anisotropy is larger for pre-SN $\overline{\nu}_{e}$'s of a few MeV than that for the SN burst $\overline{\nu}_{e}$'s of several tens of MeV, which could be illustrated in Fig.~\ref{fig:neutronCos} with the relationship of the average neutron direction and the energy of incident $\overline{\nu}_{e}$.
The average distance of neutron thermalization and diffusion before its capture on hydrogen is about 10 cm, for which the initial direction is inherited. The $\gamma$ rays emitted from the neutron capture are nearly isotropic. The energy-weighted mean position of the delayed signal is about 24 cm away from the production point. Then, the energy-deposition position of neutrons relative to that of positrons offers a possibility to determine the initial direction of $\overline{\nu}^{}_e$, and thus the direction from the detector to the SN~\cite{Vogel:1999zy}.

\begin{figure}[!t]
\begin{center}
\begin{tabular}{c}
\includegraphics[scale=0.65]{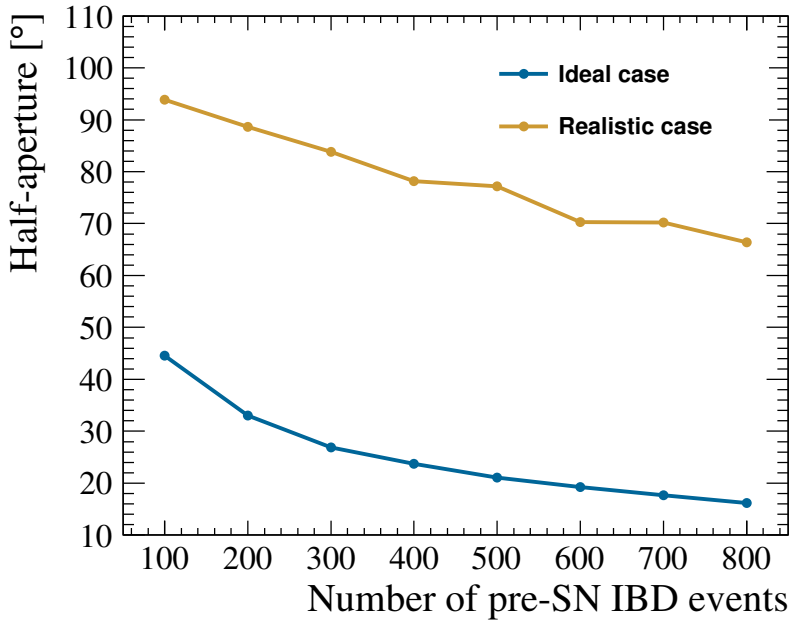}
\end{tabular}
\end{center}
\vspace{-0.7cm}
\caption{The uncertainty of the reconstructed direction for the pre-SN at the 68$\%$ confidence level for different numbers of pre-SN IBD events in the LS detector without the background from reactor IBD events. The ideal and realistic cases refer to two different scenarios of the vertex reconstruction of the prompt and delayed signals. See the main text for more details.
\label{fig:dirSig}}
\end{figure}

In practice, we reconstruct the direction of pre-SN by using the average displacement from the vertex of the prompt signal to that of the delayed signal, for which all the IBD events induced by the pre-SN $\overline{\nu}^{}_e$'s will be fully utilized. Statistically, we introduce the displacement vector $\vec{d}$, which is the reconstructed direction pointing back to the SN, namely,
\begin{eqnarray}
\vec{d} = \frac{1}{N} \sum_{i = 1}^{N} \left({\vec{x}^{i}_{e^{+}} - \vec{x}^{i}_{n}}\right) \; ,
\label{eq:dir}
\end{eqnarray}
where $N$ is the total number of IBD events from pre-SN neutrinos, $\vec{x}^i_{n}$ and $\vec{x}^i_{e^+}$ (for $i = 1, 2, \cdots, N$) denote respectively the reconstructed position vectors of neutron and position for each IBD event. We distinguish two different scenarios in which the reconstruction of $\vec{x}^{i}_{n}$ and $\vec{x}^{i}_{e^{+}}$ follows optimal and realistic assumptions, respectively. Before explaining the details of assumptions, we show the final results in the ideal and realistic scenarios in Fig.~\ref{fig:dirSig} as the blue and brown curves. For each point along the curves in Fig.~\ref{fig:dirSig}, we assume the true direction $\vec{d}^{}_{\rm t}$ of the pre-SN is along the $z$ axis, and then $\vec{d}$ is calculated with 1000 trials of the given number of simulated IBD events. The uncertainty $\sigma^{}_{\rm dir}$ of the reconstructed pre-SN direction, which is shown as the vertical axis of Fig.~\ref{fig:dirSig}, is estimated as the half-aperture of the cone around $\vec{d}^{}_{\rm t}$ which contains $68\%$ of results of 1000 trials.

\begin{itemize}
\item In \textit{the ideal case}, $\vec{x}^{i}_{n}$ refers to the position of neutron capture, while $\vec{x}^{i}_{e^{+}_{}}$ to the position of positron annihilation. The information of $\vec{x}^{i}_{n}$ and $\vec{x}^{i}_{e^{+}_{}}$ in this case have actually been obtained from the true values from each simulated IBD event. The blue curve in Fig.~\ref{fig:dirSig} shows the best results for the reconstruction of the pre-SN direction with IBD events, where one can see that the uncertainty $\sigma_{\rm dir}$ is $45^{\circ}_{}$ and $16^{\circ}_{}$ with 100 and 800 pre-SN IBD events, respectively.

\item In \textit{the realistic case}, $\vec{x}^{i}_{n}$ and $\vec{x}^{i}_{e^{+}_{}}$ refer to mean energy-weighted deposition positions of the delayed and prompt signals, respectively, which should be determined from the vertex reconstruction. For example, a useful reconstruction method has been discussed in Ref.~\cite{Liu:2018fpq}. In the realistic case, to include the resolution of the reconstructed vertex, the positions of positron annihilation and neutron capture after randomly smeared by Gaussian distributions with their variances as 10 cm and 15 cm, respectively, are taken as $\vec{x}^{i}_{n}$ and $\vec{x}^{i}_{e^{+}_{}}$. As a result, the reconstructed direction in this case is shown as the brown curve in Fig.~\ref{fig:dirSig}, where one can observe that the uncertainty $\sigma_{\rm dir}$ turns out to be $94^{\circ}_{}$ and $66^{\circ}_{}$ for 100 and 800 pre-SN IBD events, respectively. The performance in the realistic case is much worse than that in the ideal case, which can be mainly attributed to the $\gamma$ rays from the neutron capture that obscure the initial direction of neutron.

\end{itemize}
Recently, another reconstruction method has been proposed in Ref.~\cite{Wonsak:2018uby}, where it has been shown that it is possible to more precisely determine the annihilation vertex of the prompt signal. Hence, in the extreme case, $\vec{x}^{i}_{n}$ can be taken as the mean energy-weighted deposition position and $\vec{x}^{i}_{e^{+}_{}}$ as the annihilation position of the prompt positron. The uncertainty $\sigma_{\rm dir}$ of the reconstructed direction could be $90^{\circ}_{}$ and $57^{\circ}_{}$ for 100 and 800 pre-SN IBD events, respectively.

\begin{figure}[!t]
\vspace{-2cm}
\begin{center}
\begin{tabular}{l}
\includegraphics[scale=0.9]{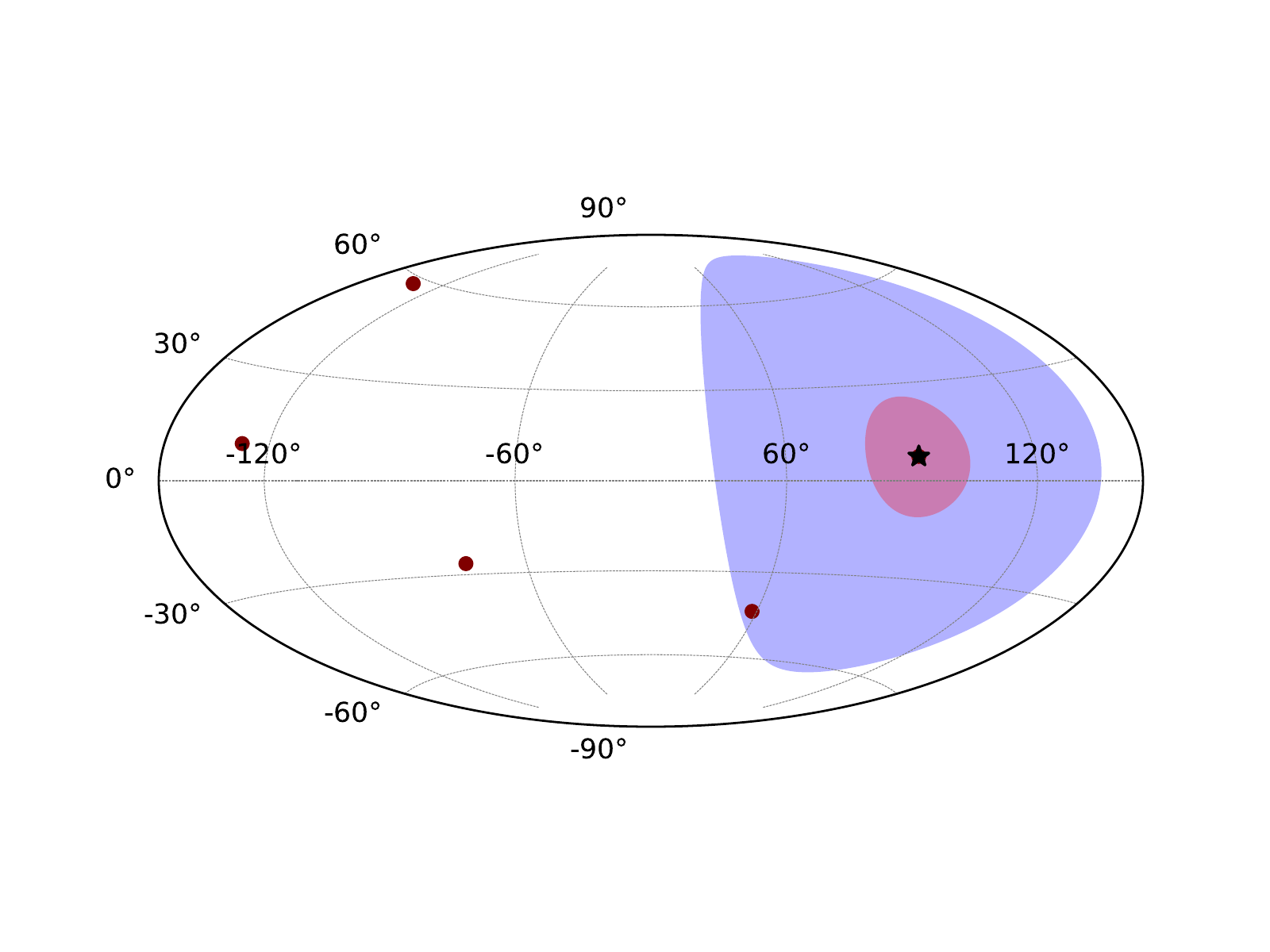}
\end{tabular}
\end{center}
\vspace{-2.5cm}
\caption{The determination of the direction of Betelgeuse (black asterisk) at $D = 0.2~{\rm kpc}$ in the equatorial system with right-ascension and declination. The mass of Betelgeuse is taken to be 25 $M^{}_{\odot}$ such that the pre-SN neutrinos from the Guo model with the same progenitor mass are assumed and neutrino oscillations in the NO case are considered. The red and blue regions refer to the ideal and realistic cases, as shown in Fig.~\ref{fig:dirSig}, respectively. The red points refer to four other RSGs within 0.34 kpc except Betelgeuse.
\label{fig:dirContour}}
\end{figure}

\begin{table}[!t]
\setlength{\tabcolsep}{11pt}
\centering
\begin{tabular}{cccccc}
\hline
\multicolumn{1}{c}{\multirow{2}{*}{Mass Ordering}}  & \multicolumn{1}{c}{} & \multicolumn{4}{c}{Pre-SN Neutrino Model} \\
\cline{3-6}\multicolumn{1}{c}{}  & \multicolumn{1}{c}{}  & \multicolumn{1}{c}{15 $M_{\odot}$ Guo} & \multicolumn{1}{c}{20 $M_{\odot}$ Guo} & \multicolumn{1}{c}{25 $M_{\odot}$ Guo} & \multicolumn{1}{c}{15 $M_{\odot}$ Kato} \\
\hline
\multicolumn{1}{l}{NO} &\multicolumn{1}{c}{}  & \multicolumn{1}{c}{83$^{\circ}$ (26$^{\circ}$)}  & \multicolumn{1}{c}{74$^{\circ}$ (20$^{\circ}$)} & \multicolumn{1}{c}{70$^{\circ}$ (19$^{\circ}$)} & \multicolumn{1}{c}{87$^{\circ}$ (31$^{\circ}$)} \\
\multicolumn{1}{l}{IO}  &\multicolumn{1}{c}{}  & \multicolumn{1}{c}{94$^{\circ}$ (45$^{\circ}$)}  & \multicolumn{1}{c}{91$^{\circ}$ (38$^{\circ}$)} & \multicolumn{1}{c}{89$^{\circ}$ (35$^{\circ}$)} & \multicolumn{1}{c}{96$^{\circ}$ (50$^{\circ}$)}  \\
\hline
\end{tabular}
\vspace{0.3cm}
\caption{The pre-SN pointing ability in the realistic case (the ideal case) for a possible SN candidate at $D = 0.2~{\rm kpc}$  using the pre-SN IBD events over the last day before collapse applied in the 20 kiloton JUNO-like LS detector, where different neutrino mass ordering and four pre-SN neutrino models with different progenitor masses and production mechanisms are considered.}
\label{table:pointing}
\end{table}
Now let us give a more concrete example for the SN pointing. Assuming pre-SN $\overline{\nu}_{e}$'s from a progenitor star at the distance of $D = 0.2~{\rm kpc}$ (e.g., similar to Betelgeuse), we illustrate the directionality accuracy of the 20 kiloton JUNO-like detector in Table~\ref{table:pointing} by considering different neutrino mass ordering and four pre-SN neutrino models with different progenitor masses and production mechanisms. For the NO case, the direction of the progenitor star can be localized within $19^{\circ}$, $20^{\circ}$, $26^{\circ}$ and $31^{\circ}$ in the ideal case and $70^{\circ}_{}$, $74^{\circ}$, $83^{\circ}$ and $87^{\circ}$ in the realistic case at the $68\%$ confidence level, respectively. Meanwhile the direction accuracy would be worse for the IO case because of the relatively lower IBD statistics.
In Fig.~\ref{fig:dirContour}, we take the 25 $M_{\odot}$ Guo model and NO case as an example to show these two uncertainties as red and blue regions in the equatorial system with right-ascension and declination, where the location of the true progenitor star is represented by the black asterisk.
For such a setup, we have about 650 IBD events within the last one day before collapse after applying the energy cut of $0.9~{\rm MeV} \le E^{}_{\rm o} \le 3.5~{\rm MeV}$ and the detection efficiency of $73\%$, and the direction of the progenitor star can be localized within $19^{\circ}$ in the ideal case and $70^{\circ}_{}$ in the realistic case at the $68\%$ confidence level. The limited number of known SN candidates in our galaxy as shown in Fig.~\ref{fig:sigRSG} could help to further narrow down the region of SN candidates. Four other RSGs within 0.34 kpc represented by the red points in Fig.~\ref{fig:dirContour} are shown for illustration. For the next nearby galactic SN, the pre-SN direction could be reconstructed as long as enough pre-SN IBD events will be accumulated. In this case, the pre-SN early warning accompanied with its direction could be provided from future large LS detectors, which are extremely important for the astronomical communities.

The SN early warning before the burst may also be achieved by monitoring the variability of progenitors with the wide field telescopes~\cite{Szczygiel:2011de,Adams:2013ana,Kochanek:2016cvd,Shappee:2013mna,Holoien:2018kcp}, which could cover a rather broadband wavelength and long distance. However, since it is still not well-established on the rate of magnitude variability~\cite{Szczygiel:2011de,Kochanek:2016cvd,Johnson:2017hcj,Hillier:2019jvx} and light characteristics~\cite{Nakamura:2016kkl} during the late stage of the SN progenitors, independent early warning for the nearby galactic SN using pre-SN neutrinos is still useful and complementary to the optical telescope monitoring. The early warning and possible directional information will be helpful for the telescopes to catch the early-time light from the SN.

\section{Summary} \label{sec:IV}

In this paper, we have studied prospects of the future large LS detectors for the pre-SN neutrino observation. For a JUNO-like detector~\cite{An:2015jdp} and a set of stellar models with different progenitor star masses~\cite{Guo:2019orq, Kato:2017ehj}, both the IBD and $e$ES event spectra have been calculated. We emphasize that the total number of the $e$ES events may be even larger than that of the IBD events. However, as the $e$ES events in the LS detector are more concentrated in the low-energy region of $E^{}_{\rm o} \lesssim 2~{\rm MeV}$ and show up as singles, it is quite challenging to identify them unless the backgrounds from the intrinsic radioactivity of LS, detector materials and cosmogenic isotopes can be significantly reduced. If unambiguously observed, the $e$ES events could provide us with complementary information on the pre-SN neutrinos of all flavors, while the IBD events are induced only by electron antineutrinos. This is indispensable for the understanding of neutrino production mechanism and the theory of late-time evolution of massive stars.

The LS detector with a low threshold of observable energies (e.g., $E^{\rm th}_{\rm o} = 0.2~{\rm MeV}$ for JUNO) is particularly advantageous and important for the detection of pre-SN neutrinos, whose energies are expected to be a few MeV. Taking the JUNO detector as described in Ref.~\cite{An:2015jdp} for example, we focus only on the IBD events due to their clean coincidence signals and attempt to answer three questions related to the early warning of the core-collapse SNe.
\begin{enumerate}
\item {\it How far the SN distance can be reached}. The answer crucially depends on the stellar models of the progenitor stars and neutrino flavor conversions in either NO or IO case. In the most optimal scenario (with the progenitor mass of $25~M^{}_{\odot}$ and NO) and the worst-case scenario (with the progenitor mass of $12~M^{}_{\odot}$ and IO), the SN distance that can be reached with a $3\sigma$ significance of the detection of pre-SN neutrinos turns out to be $1.23~{\rm kpc}$ and $0.34~{\rm kpc}$, respectively, which covers 45 RSG stars and 5 RSG stars according to the collection of RSG stars from Ref.~\cite{Nakamura:2016kkl}.

\item {\it How long before the collapse the warning can be sent out}. Again the answer is largely dependent on the stellar models and neutrino mass ordering. To be specific, we assume that the pre-SN star is located at the distance of $D = 0.2~{\rm kpc}$ (i.e., the distance of the SN candidate, Betelgeuse) and the Guo model~\cite{Guo:2019orq} with a progenitor mass of $15~M^{}_\odot$. In this case, the warning of a core-collapse SN via the detection of pre-SN neutrinos with a $3\sigma$ significance can be sent out by 3.2 days (or 0.3 days) prior to collapse if NO (or IO) is assumed. Even in the latter case, the early warning of a core-collapse SN can be made about eight hours in advance.

\item {\it How well the SN direction can be determined.} Following the idea suggested in Ref.~\cite{Vogel:1999zy}, we carry out a statistical determination of the SN direction by using the displacement of final-state neutrons and positrons from all the IBD events of pre-SN neutrinos. For Betelgeuse at $D = 0.2~{\rm kpc}$, we adopt the stellar model of a $25~M^{}_{\odot}$ progenitor mass and obtain about $650$ IBD events in total after taking into account the energy cut and the detection efficiency. In the realistic case, where the positions of the neutron and the positron are obtained from the vertex reconstruction, the uncertainty of the reconstructed SN direction is found to be $70^\circ$ at the $68\%$ confidence level. As there are just five red supergiant stars, which may ultimately end their lives as core-collapse SNe, within a distance of $0.34~{\rm kpc}$ in our galaxy, such an accuracy in the determination of the SN direction is already very informative.
\end{enumerate}

It is also worthwhile to mention that the large dark-matter detectors could also record pre-SN neutrinos via the coherent neutrino-nucleus scattering~\cite{Raj:2019wpy}. Given a rather simple picture of flavor conversions for the pre-SN neutrinos, the luminosities, average energies and flavor content of the original pre-SN neutrinos could be unveiled with a good opportunity after combining different reaction channels in the neutrino detectors and even the pre-SN neutrino events in the dark-matter detectors. All these efforts will finally lead to more insights into the theory of the late-time evolution of massive stars.

\section*{ACKNOWLEDGEMENTS}

The authors thank G. Guo, Y. Z. Qian and C. Kato for kindly providing the neutrino data from their stellar models. This work was supported in part by the National Key R$\&$D Program of China under Grant No. 2018YFA0404100, by the Strategic Priority Research Program of the Chinese Academy of Sciences under Grant No. XDA10010100,  by the National Natural Science Foundation of China under Grant No. 11775232 and No. 11835013, by the CAS Center for Excellence in Particle Physics (CCEPP) and by the China Postdoctoral Science Foundation funded project under Grant No. 2019M650844.



\begin{thebibliography}{99}

\bibitem{Woosley:2002zz}
  S.~E.~Woosley, A.~Heger and T.~A.~Weaver,
  ``The evolution and explosion of massive stars,''
  Rev.\ Mod.\ Phys.\  {\bf 74}, 1015 (2002).

\bibitem{Janka:2017vlw}
  H.-T.~Janka,
  ``Neutrino Emission from Supernovae,''
 in {\it Handbook of Supernovae}, edited by A.~Alsabti and P.~Murdin, p. 1575, Springer International Publishing AG, 2017.

\bibitem{Hirata:1987hu}
  K.~Hirata {\it et al.} [Kamiokande-II Collaboration],
  ``Observation of a Neutrino Burst from the Supernova SN 1987a,''
  Phys.\ Rev.\ Lett.\  {\bf 58}, 1490 (1987).

\bibitem{Bionta:1987qt}
  R.~M.~Bionta {\it et al.},
  ``Observation of a Neutrino Burst in Coincidence with Supernova SN 1987a in the Large Magellanic Cloud,''
  Phys.\ Rev.\ Lett.\  {\bf 58}, 1494 (1987).

\bibitem{Alekseev:1988gp}
  E.~N.~Alekseev, L.~N.~Alekseeva, I.~V.~Krivosheina and V.~I.~Volchenko,
  ``Detection of the Neutrino Signal From {SN1987A} in the {LMC} Using the Inr Baksan Underground Scintillation Telescope,''
  Phys.\ Lett.\ B {\bf 205}, 209 (1988).

\bibitem{Odrzywolek:2003vn}
  A.~Odrzywolek, M.~Misiaszek and M.~Kutschera,
  ``Detection possibility of the pair - annihilation neutrinos from the neutrino - cooled pre-supernova star,''
  Astropart.\ Phys.\  {\bf 21}, 303 (2004)

\bibitem{Odrzywolek:2010zz}
  A.~Odrzywolek and A.~Heger,
  ``Neutrino signatures of dying massive stars: From main sequence to the neutron star,''
  Acta Phys.\ Polon.\ B {\bf 41}, 1611 (2010).

\bibitem{Beaudet:1967zz}
  G.~Beaudet, V.~Petrosian and E.~E.~Salpeter,
  ``Energy Losses due to Neutrino Processes,''
  Astrophys.\ J.\  {\bf 150}, 979 (1967).

\bibitem{Munakata1985}
  H.~Munakata, Y.~Kohyama and N.~G.~Itoh,
  ``Neutrino Energy Loss in Stellar Interiors,''
  Astrophys.\ J.\  {\bf 296}, 197 (1985).

\bibitem{Schinder:1986nh}
  P.~J.~Schinder, D.~N.~Schramm, P.~J.~Wiita, S.~H.~Margolis and D.~L.~Tubbs,
  ``Neutrino Emission by the Pair, Plasma and Photoprocesses in the Weinberg-Salam Model,''
  Astrophys.\ J.\  {\bf 313}, 531 (1987).

\bibitem{Itoh:1989}
N.~Itoh, T.~Adachi, M.~Nakagawa, Y.~Kohyama and H.~Munakata,
``Neutrino Energy Loss in Stellar Interiors. III. Pair, Photo-, Plasma, and Bremsstrahlung Processes,"
  Astrophys.\ J.\  {\bf 339}, 354 (1989).

\bibitem{Haft:1993jt}
  M.~Haft, G.~Raffelt and A.~Weiss,
  ``Standard and nonstandard plasma neutrino emission revisited,''
  Astrophys.\ J.\  {\bf 425}, 222 (1994)
  Erratum: [Astrophys.\ J.\  {\bf 438}, 1017 (1995)]

\bibitem{Itoh:1996}
N. Itoh, H. Hayashi, A. Nishikawa, and Y. Kohyama,
``Neutrino Energy Loss in Stellar Interiors. VII. Pair, Photo-, Plasma, Bremsstrahlung, and Recombination Neutrino Processes,"
Astrophys. J. Suppl. Ser. {\bf 102}, 411 (1996).

\bibitem{Guo:2016vls}
  G.~Guo and Y.~Z.~Qian,
  ``Spectra and rates of bremsstrahlung neutrino emission in stars,''
  Phys.\ Rev.\ D {\bf 94}, no. 4, 043005 (2016)
  [arXiv:1608.02852].

\bibitem{Yoshida:2016imf}
  T.~Yoshida, K.~Takahashi, H.~Umeda and K.~Ishidoshiro,
  ``Presupernova neutrino events relating to the final evolution of massive stars,''
  Phys.\ Rev.\ D {\bf 93}, no. 12, 123012 (2016)
  [arXiv:1606.04915].

\bibitem{Kato:2017ehj}
  C.~Kato, H.~Nagakura, S.~Furusawa, K.~Takahashi, H.~Umeda, T.~Yoshida, K.~Ishidoshiro and S.~Yamada,
  ``Neutrino emissions in all flavors up to the pre-bounce of massive stars and the possibility of their detections,''
  Astrophys.\ J.\  {\bf 848}, no. 1, 48 (2017)
  [arXiv:1704.05480].

\bibitem{Patton:2017neq}
  K.~M.~Patton, C.~Lunardini, R.~J.~Farmer and F.~X.~Timmes,
  ``Neutrinos from beta processes in a presupernova: probing the isotopic evolution of a massive star,''
  Astrophys.\ J.\  {\bf 851}, no. 1, 6 (2017)
  [arXiv:1709.01877].

\bibitem{Fukuda:2002uc}
  Y.~Fukuda {\it et al.} [Super-Kamiokande Collaboration],
  ``The Super-Kamiokande detector,''
  Nucl.\ Instrum.\ Meth.\ A {\bf 501}, 418 (2003).

\bibitem{Abe:2016nxk}
  K.~Abe {\it et al.} [Super-Kamiokande Collaboration],
  ``Solar Neutrino Measurements in Super-Kamiokande-IV,''
  Phys.\ Rev.\ D {\bf 94}, no. 5, 052010 (2016)
  [arXiv:1606.07538].

\bibitem{Abe:2011ts}
  K.~Abe {\it et al.},
  ``Letter of Intent: The Hyper-Kamiokande Experiment --- Detector Design and Physics Potential ---,''
  [arXiv:1109.3262].

\bibitem{Abe:2018uyc}
  K.~Abe {\it et al.} [Hyper-Kamiokande Collaboration],
  ``Hyper-Kamiokande Design Report,''
  [arXiv:1805.04163].

\bibitem{Asakura:2015bga}
  K.~Asakura {\it et al.} [KamLAND Collaboration],
  ``KamLAND Sensitivity to Neutrinos from Pre-Supernova Stars,''
  Astrophys.\ J.\  {\bf 818}, no. 1, 91 (2016)
  
  [arXiv:1506.01175].

\bibitem{Cadonati:2000kq}
  L.~Cadonati, F.~P.~Calaprice and M.~C.~Chen,
  ``Supernova neutrino detection in borexino,''
  Astropart.\ Phys.\  {\bf 16}, 361 (2002)


\bibitem{Mukhopadhyay:2020ubs}
M.~Mukhopadhyay, C.~Lunardini, F.~Timmes and K.~Zuber,
``Presupernova neutrinos: directional sensitivity and prospects for progenitor identification,''
[arXiv:2004.02045].

\bibitem{An:2015jdp}
  F.~An {\it et al.} [JUNO Collaboration],
  ``Neutrino Physics with JUNO,''
  J.\ Phys.\ G {\bf 43}, no. 3, 030401 (2016)
  [arXiv:1507.05613].

\bibitem{Wurm:2011zn}
  M.~Wurm {\it et al.} [LENA Collaboration],
  ``The next-generation liquid-scintillator neutrino observatory LENA,''
  Astropart.\ Phys.\  {\bf 35}, 685 (2012)
  [arXiv:1104.5620].

\bibitem{Simpson:2019xwo}
  C.~Simpson {\it et al.} [Super-Kamiokande Collaboration],
  ``Sensitivity of Super-Kamiokande with Gadolinium to Low Energy Anti-neutrinos from Pre-supernova Emission,''
  arXiv:1908.07551.

\bibitem{Acciarri:2016crz}
  R.~Acciarri {\it et al.} [DUNE Collaboration],
  ``Long-Baseline Neutrino Facility (LBNF) and Deep Underground Neutrino Experiment (DUNE) : Conceptual Design Report, Volume 1: The LBNF and DUNE Projects,''
  arXiv:1601.05471.

\bibitem{Guo:2019orq}
  G.~Guo, Y.~Z.~Qian and A.~Heger,
  ``Presupernova neutrino signals as potential probes of neutrino mass hierarchy,''
  Phys.\ Lett.\ B {\bf 796}, 126 (2019)
  [arXiv:1906.06839].

\bibitem{Pantaleone:1992eq}
  J.~T.~Pantaleone,
  ``Neutrino oscillations at high densities,''
  Phys.\ Lett.\ B {\bf 287}, 128 (1992).

\bibitem{Samuel:1993uw}
  S.~Samuel,
  ``Neutrino oscillations in dense neutrino gases,''
  Phys.\ Rev.\ D {\bf 48}, 1462 (1993).

\bibitem{Duan:2010bg}
  H.~Duan, G.~M.~Fuller and Y.~Z.~Qian,
  ``Collective Neutrino Oscillations,''
  Ann.\ Rev.\ Nucl.\ Part.\ Sci.\  {\bf 60}, 569 (2010)
  [arXiv:1001.2799].

\bibitem{Chakraborty:2016yeg}
  S.~Chakraborty, R.~Hansen, I.~Izaguirre and G.~Raffelt,
  ``Collective neutrino flavor conversion: Recent developments,''
  Nucl.\ Phys.\ B {\bf 908}, 366 (2016)
  [arXiv:1602.02766].

\bibitem{Mirizzi:2015eza}
  A.~Mirizzi, I.~Tamborra, H.~T.~Janka, N.~Saviano, K.~Scholberg, R.~Bollig, L.~Hudepohl and S.~Chakraborty,
  ``Supernova Neutrinos: Production, Oscillations and Detection,''
  Riv.\ Nuovo Cim.\  {\bf 39}, no. 1-2, 1 (2016)
  [arXiv:1508.00785].



\bibitem{Wolfenstein:1977ue}
  L.~Wolfenstein,
  ``Neutrino Oscillations in Matter,''
  Phys.\ Rev.\ D {\bf 17}, 2369 (1978).

\bibitem{Mikheev:1986gs}
  S.~P.~Mikheev and A.~Y.~Smirnov,
  ``Resonance Amplification of Oscillations in Matter and Spectroscopy of Solar Neutrinos,''
  Sov.\ J.\ Nucl.\ Phys.\  {\bf 42}, 913 (1985)
  [Yad.\ Fiz.\  {\bf 42}, 1441 (1985)].

\bibitem{Dighe:1999bi}
  A.~S.~Dighe and A.~Y.~Smirnov,
  ``Identifying the neutrino mass spectrum from the neutrino burst from a supernova,''
  Phys.\ Rev.\ D {\bf 62}, 033007 (2000)

\bibitem{Tanabashi:2018oca}
  M.~Tanabashi {\it et al.} [Particle Data Group],
  ``Review of Particle Physics,''
  Phys.\ Rev.\ D {\bf 98}, no. 3, 030001 (2018).

\bibitem{Dighe:2003jg}
  A.~S.~Dighe, M.~T.~Keil and G.~G.~Raffelt,
  ``Identifying earth matter effects on supernova neutrinos at a single detector,''
  JCAP {\bf 0306}, 006 (2003)
  [hep-ph/0304150].

\bibitem{Mirizzi:2006xx}
  A.~Mirizzi, G.~G.~Raffelt and P.~D.~Serpico,
  ``Earth matter effects in supernova neutrinos: Optimal detector locations,''
  JCAP {\bf 0605}, 012 (2006)
  [astro-ph/0604300].

\bibitem{Borriello:2012zc}
  E.~Borriello, S.~Chakraborty, A.~Mirizzi, P.~D.~Serpico and I.~Tamborra,
  ``(Down-to-)Earth matter effect in supernova neutrinos,''
  Phys.\ Rev.\ D {\bf 86}, 083004 (2012)
  [arXiv:1207.5049].

\bibitem{Liao:2016uis}
  W.~Liao,
  ``Detecting supernovae neutrino with Earth matter effect,''
  Phys.\ Rev.\ D {\bf 94}, no. 11, 113016 (2016)
  [arXiv:1607.03334].


\bibitem{Abe:2016nxk}
  K.~Abe {\it et al.} [Super-Kamiokande Collaboration],
  Phys.\ Rev.\ D {\bf 94}, no. 5, 052010 (2016)
  [arXiv:1606.07538].

\bibitem{Lujan-Peschard:2014lta}
  C.~Lujan-Peschard, G.~Pagliaroli and F.~Vissani,
  ``Spectrum of Supernova Neutrinos in Ultra-pure Scintillators,''
  JCAP {\bf 1407}, 051 (2014)
  [arXiv:1402.6953].

\bibitem{Lu:2016ipr}
  J.~S.~Lu, Y.~F.~Li and S.~Zhou,
  ``Getting the most from the detection of Galactic supernova neutrinos in future large liquid-scintillator detectors,''
  Phys.\ Rev.\ D {\bf 94}, no. 2, 023006 (2016)
  [arXiv:1605.07803].

\bibitem{Li:2017dbg}
  H.~L.~Li, Y.~F.~Li, M.~Wang, L.~J.~Wen and S.~Zhou,
  ``Towards a complete reconstruction of supernova neutrino spectra in future large liquid-scintillator detectors,''
  Phys.\ Rev.\ D {\bf 97}, no. 6, 063014 (2018)
  [arXiv:1712.06985].



 \bibitem{Harper:2008}
 G. M.~Harper, A.~Brown, and E. F.~Guinan,
 ``A new VLA-Hipparcos distance to Betelgeuse and its implications,''
 Astron. J. {\bf 135}, 1430 (2008).

\bibitem{Smith:2008ef}
  N.~Smith, K.~H.~Hinkle and N.~Ryde,
  ``Red supergiants as potential Type IIn supernova progenitors: Spatially resolved 4.6 micron CO emission around VY CMa and Betelgeuse,''
  Astron.\ J.\  {\bf 137}, 3558 (2009)
  [arXiv:0811.3037].

\bibitem{Neilson:2011ta}
  H.~Neilson, J.~B.~Lester and X.~Haubois,
  ``Weighing Betelgeuse: Measuring the mass of alpha Orionis from stellar limb-darkening,''
  ASP Conf.\ Ser.\  {\bf 451}, 117 (2011)
  [arXiv:1109.4562].

\bibitem{Dolan:2017}
  M.~Dolan, G. ~Mathews, D.~Lam, N. Q.~Lan, G. J. Herczeg, D.~Dearborn.
  ``Evolutionary tracks for Betelgeuse,''
  Astron.\ J.\  {\bf 819}, 7(2017)
  [arXiv:1406.3143]


\bibitem{Cowan:1998}
G. D.~Cowan,
``Statistical Data Analysis,''
Oxford University Press (1998).


\bibitem{Antonioli:2004zb}
  P.~Antonioli {\it et al.},
  ``SNEWS: The Supernova Early Warning System,''
  New J.\ Phys.\  {\bf 6}, 114 (2004)


\bibitem{Strumia:2003zx}
A.~Strumia and F.~Vissani,
``Precise quasielastic neutrino/nucleon cross-section,''
Phys. Lett. B \textbf{564}, 42-54 (2003)


\bibitem{Vogel:1999zy}
  P.~Vogel and J.~F.~Beacom,
  ``Angular distribution of neutron inverse beta decay, $\overline{\nu}^{}_e + p \to e^+ + n$,''
  Phys.\ Rev.\ D {\bf 60}, 053003 (1999)

\bibitem{Apollonio:1999jg}
  M.~Apollonio {\it et al.} [CHOOZ Collaboration],
  ``Determination of neutrino incoming direction in the CHOOZ experiment and supernova explosion location by scintillator detectors,''
  Phys.\ Rev.\ D {\bf 61}, 012001 (2000)


\bibitem{Fischer:2015oma}
V.~Fischer, T.~Chirac, T.~Lasserre, C.~Volpe, M.~Cribier, M.~Durero, J.~Gaffiot, T.~Houdy, A.~Letourneau, G.~Mention, M.~Pequignot, V.~Sibille and M.~Vivier,
``Prompt directional detection of galactic supernova by combining large liquid scintillator neutrino detectors,''
JCAP \textbf{08}, 032 (2015)


\bibitem{Liu:2018fpq}
  Q.~Liu, M.~He, X.~Ding, W.~Li and H.~Peng,
  ``A vertex reconstruction algorithm in the central detector of JUNO,''
  JINST {\bf 13}, no. 09, T09005 (2018)
  [arXiv:1803.09394].

\bibitem{Wonsak:2018uby}
  B.~S.~Wonsak {\it et al.},
  ``Topological track reconstruction in unsegmented, large-volume liquid scintillator detectors,''
  JINST {\bf 13}, no. 07, P07005 (2018)
  [arXiv:1803.08802].

\bibitem{Adams:2013ana}
  S.~M.~Adams, C.~S.~Kochanek, J.~F.~Beacom, M.~R.~Vagins and K.~Z.~Stanek,
  ``Observing the Next Galactic Supernova,''
  Astrophys.\ J.\  {\bf 778}, 164 (2013)
  [arXiv:1306.0559].

\bibitem{Shappee:2013mna}
  B.~J.~Shappee {\it et al.},
  ``The Man Behind the Curtain: X-rays Drive the UV through NIR Variability in the 2013 AGN Outburst in NGC 2617,''
  Astrophys.\ J.\  {\bf 788}, 48 (2014)
  [arXiv:1310.2241].

\bibitem{Holoien:2018kcp}
  T.~W.-S.~Holoien {\it et al.},
  ``The ASAS-SN bright supernova catalogue - IV. 2017,''
  Mon.\ Not.\ Roy.\ Astron.\ Soc.\  {\bf 484}, no. 2, 1899 (2019)
  [arXiv:1811.08904].

\bibitem{Szczygiel:2011de}
  D.~M.~Szczygiel, J.~R.~Gerke, C.~S.~Kochanek and K.~Z.~Stanek,
  ``Discovery of Variability of the Progenitor of SN 2011dh in M51 Using the Large Binocular Telescope,''
  Astrophys.\ J.\  {\bf 747}, 23 (2012)
  [arXiv:1110.2783].

\bibitem{Kochanek:2016cvd}
  C.~S.~Kochanek {\it et al.},
  ``Supernova Progenitors, Their Variability, and the Type IIP Supernova ASASSN-16fq in M66,''
  Mon.\ Not.\ Roy.\ Astron.\ Soc.\  {\bf 467}, no. 3, 3347 (2017)
  [arXiv:1609.00022].

\bibitem{Johnson:2017hcj}
  S.~A.~Johnson, C.~S.~Kochanek and S.~M.~Adams,
  ``The quiescent progenitors of four Type II-P/L supernovae,''
  Mon.\ Not.\ Roy.\ Astron.\ Soc.\  {\bf 480}, no. 2, 1696 (2018)
  [arXiv:1712.03957].

\bibitem{Hillier:2019jvx}
  D.~J.~Hillier and L.~Dessart,
  ``On the photometric and spectroscopic diversity of Type II supernovae,''
  Astron.\ Astrophys.\  {\bf 631}, A8 (2019)
  [arXiv:1908.02973].

\bibitem{Nakamura:2016kkl}
  K.~Nakamura, S.~Horiuchi, M.~Tanaka, K.~Hayama, T.~Takiwaki and K.~Kotake,
  ``Multimessenger signals of long-term core-collapse supernova simulations: synergetic observation strategies,''
  Mon.\ Not.\ Roy.\ Astron.\ Soc.\  {\bf 461}, no. 3, 3296 (2016)
  [arXiv:1602.03028].

\bibitem{Raj:2019wpy}
  N.~Raj, V.~Takhistov and S.~J.~Witte,
  ``Pre-Supernova Neutrinos in Large Dark Matter Direct Detection Experiments,''
  [arXiv:1905.09283].

\end{thebibliography}
\end{document}